\documentclass[10pt, conference, letterpaper]{IEEEtran}
\IEEEoverridecommandlockouts

\usepackage{booktabs} 
\usepackage[noadjust]{cite}
\usepackage{hyperref} 
\usepackage{amsmath}
\usepackage{amssymb}
\usepackage{amsfonts}
\usepackage{float}
\usepackage{graphicx}
\usepackage{pifont}
\usepackage{colortbl}
\usepackage{multirow}
\usepackage{mathtools}
\usepackage{algorithm}
\usepackage{algorithmicx}
\usepackage{algpseudocode}
\usepackage{tabto}
\usepackage{hhline}
\usepackage{enumitem}
\usepackage{balance}
\usepackage[font=small,labelfont={it}]{caption}
\usepackage{xargs}

\usepackage{fancyhdr}
\usepackage{tikz}

\makeatletter
\renewcommand\footnoterule{%
  \kern-3\p@
  \hrule\@width 0.5\columnwidth
  \kern2.6\p@}
 \makeatother

\usepackage{soul}

\usepackage[colorinlistoftodos,prependcaption,textsize=tiny]{todonotes}
\newcommandx{\shafique}[2][1=]{\todo[linecolor=red,backgroundcolor=red!25,bordercolor=red,#1]{#2}}

\begin{document}
%
%
\title{Robust Machine Learning Systems: Challenges, Current Trends, Perspectives, and the Road Ahead}
\author{\IEEEauthorblockN {Muhammad Shafique$^1$, Mahum Naseer$^1$, Theocharis Theocharides$^2$, Christos Kyrkou$^2$, \\Onur Mutlu$^3$, Lois  Orosa$^3$, Jungwook Choi$^4$}
  	\IEEEauthorblockA{$^1$Technische Universit\"at Wien (TU Wien), Vienna, Austria,\\
  	$^2$University of Cyprus, Cyprus\\
  	$^3$ETH Z\"urich, Z\"urich, Switzerland,\\
  	$^4$Hanyang University, South Korea}
  	
  	\IEEEauthorblockA{Corresponding Authors' Email: muhammad.shafique@tuwien.ac.at, ttheocharides@ucy.ac.cy}
  	}
\maketitle
\begin{abstract}

Machine Learning (ML) techniques have been rapidly adopted by smart Cyber-Physical Systems (CPS) and Internet-of-Things (IoT) due to their powerful decision-making capabilities. However, they are vulnerable to various security and reliability threats, at both hardware and software levels, that compromise their accuracy. These threats get aggravated in emerging edge ML devices that have stringent constraints in terms of resources (e.g., compute,  memory, power/energy), and that therefore cannot employ costly security and reliability measures. Security, reliability, and vulnerability mitigation techniques span from network security measures to hardware protection, with an increased interest towards formal verification of trained ML models.

This paper summarizes the prominent vulnerabilities of modern ML systems, highlights successful defenses and mitigation techniques against these vulnerabilities, both at the cloud (i.e., during the ML training phase) and edge (i.e., during the ML inference stage), discusses the implications of a resource-constrained design on the reliability and security of the system, identifies verification methodologies to ensure correct system behavior, and describes open research challenges for building secure and reliable ML systems at both the edge and the cloud.
\end{abstract}


\section{Introduction} \label{intro}

Fueled by independent developments in semiconductor technology, computing, communication, control signal generation, sensors and actuators, the concept of a unified Smart Cyber-Physical System (CPS) has evolved into a ubiquitous paradigm. CPS, as the name implies, links the cyber and the physical environments with smart control. 
Together with the evolution of Internet-of-Things (IoT), which provides remote access to the CPS for controlling and monitoring the inter-connected computing devices, the standard architecture of a Smart CPS comprises three major layers~\cite{cps-architecture}: edge, fog and cloud. 
The \textit{edge} of the system is what connects the system to the physical environment, for instance, the sensors. 
The \textit{fog} is the central layer where most system computations generally occur. However, to reduce transmission overhead or for data privacy, initial computations may occur at the edge too. 
The \textit{cloud} is what connects the system to a large-scale cyberspace, which performs extensive processing, storage and communication between different cyber-physical systems. 

Improving the decision making, monitoring and control capabilities across different CPS/IoT layers is critical for emerging applications.
As the complexity, volume, and rate of data produced by IoT with many smart cyber-physical systems is increasing, Machine Learning (ML) has emerged as a dominant paradigm for analytics, decision-making, perception, and understanding.  Consequently, reliability and security vulnerabilities of ML systems can have cascading effects on smart CPS applications and critically impact ML operation across all layers.

\begin{figure*}[ht]
	\centering
	\includegraphics[width=\linewidth]{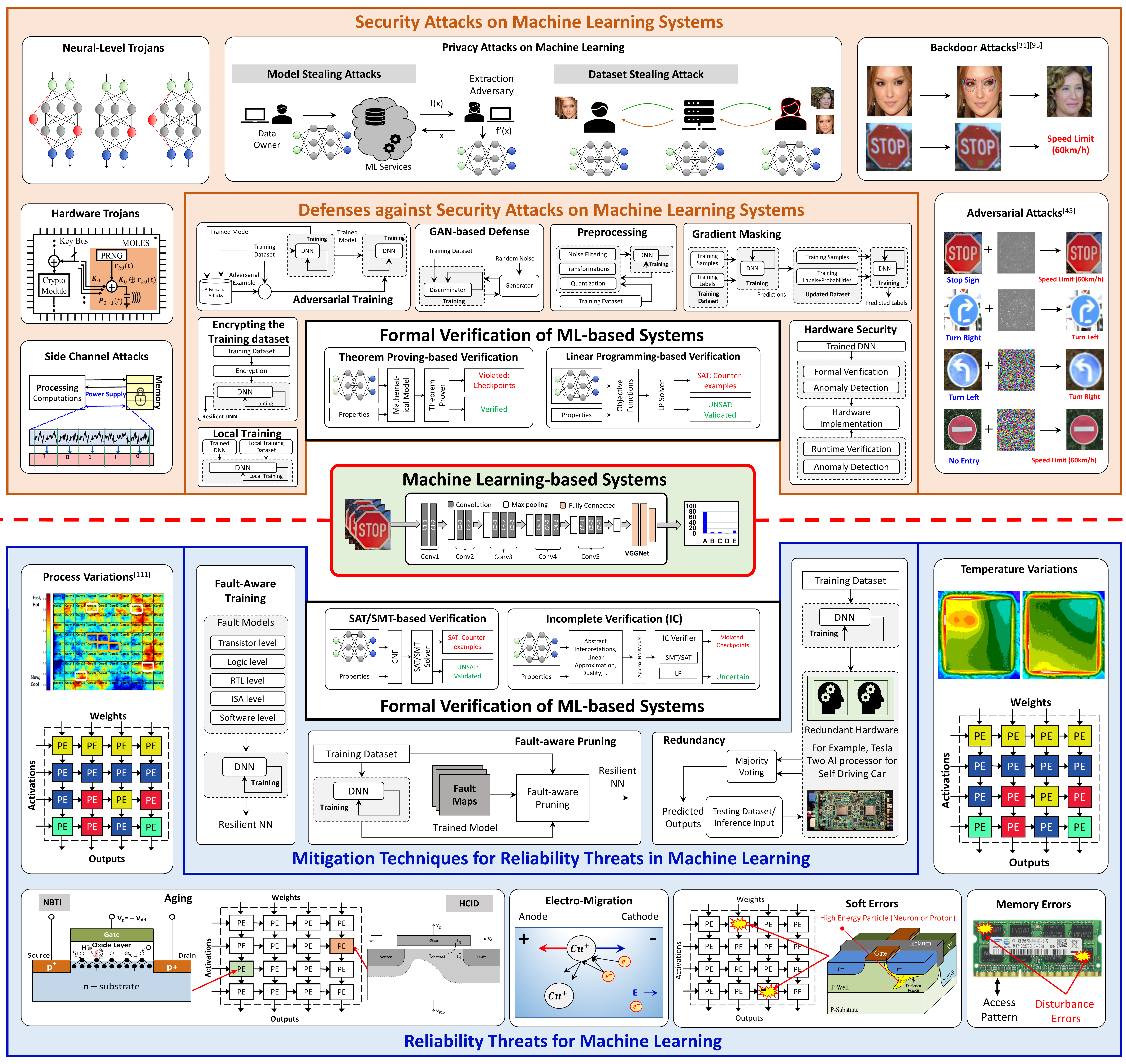}
	\caption{\textit{Overview of threats and challenges associated with ML-based systems: reliability threats and corresponding mitigation techniques (bottom), and security attacks and corresponding defenses (top).}}
	\label{fig:challenges}
\end{figure*}

\begin{figure*}[ht]
	\centering
	\includegraphics[width=\linewidth]{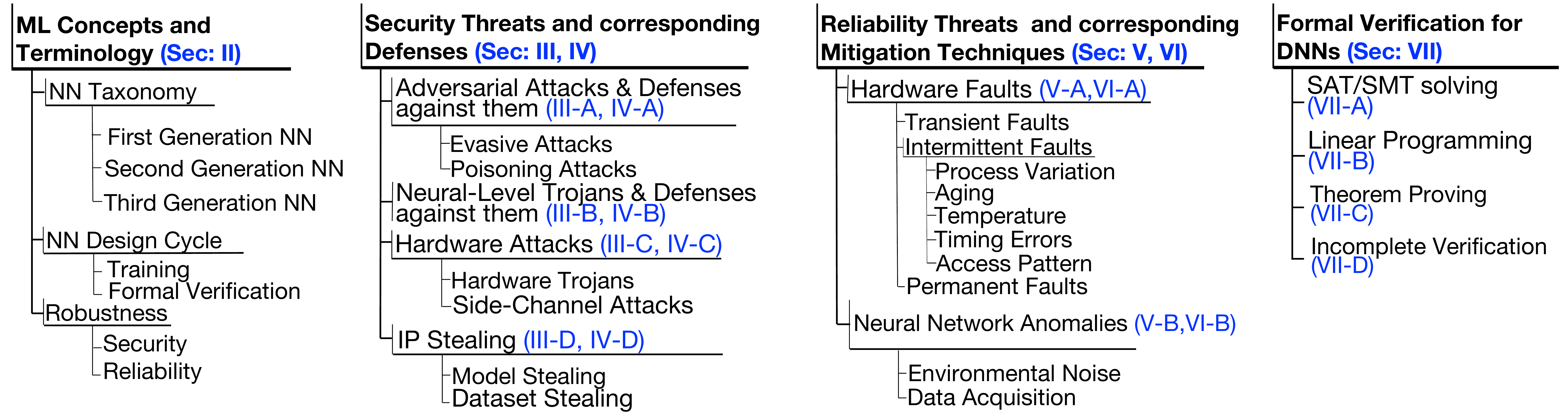}
	\caption{\textit{Organization of the paper.}}
	\label{po}
\end{figure*}

The most recent developments in Machine Learning (ML), especially in Deep Learning, evolved from the concept of a single-layer neural network, the perceptron \cite{rosenblatt1957}, to the Multi-Layer Perceptron (MLP)~\cite{rosenblatt1957} and the current intricate multi-layer Deep Neural Networks (DNNs) \cite{vggnet}\cite{densenet}, with the objective of approaching and even exceeding human decision making capabilities for a certain set of tasks. 
Due to their effectiveness at handling large amounts of data, learning (and sometimes re-learning \cite{L2Mdarpa}) input characteristics, and demonstrating high accuracy during inference of unseen inputs, ML systems have proliferated to numerous real-world applications. These include object detection \cite{auto-drive1}, face recognition \cite{face}, speech recognition \cite{speech}, spam filtering \cite{spam}, malware detection \cite{android-malware}, smart grids \cite{smart-grid}, and even safety-critical applications like autonomous driving \cite{auto-drive2}, intelligent transportation \cite{ITS} and health-care \cite{health1}\cite{health2}, where errors may lead to catastrophic results. 

Despite their high inference accuracy in practical applications, ML systems are highly vulnerable to security and reliability threats at both the cloud and the edge. Poisoning the training data (e.g., by inserting random or crafted noise to the data) with incorrectly-labelled inputs, inserting malicious components into the system hardware, polluting inputs with imperceptible noise during inference (i.e., during the run-time operation of a system), and monitoring system side channels to deduce the underlying model are some of the ways in which an attacker can breach the security of an ML system. 
Even in the absence of an explicit attacker, process variation during hardware fabrication, memory errors, environmental conditions around the system during training and inference can compromise the reliability of an ML system. Approaches to defend ML systems against these concerns exist, but each approach has its own limitations. Fig. \ref{fig:challenges} summarizes both the security and reliability threats that can affect the accuracy of ML systems and their respective countermeasures.

In this article, we aim to provide a comprehensive overview of vulnerabilities that affect modern ML systems, survey  state-of-the-art attacks and defense mechanisms, describe different solution directions and challenges, and identify potential promising avenues to research.

To ease reading, we provide the list of the acronyms used in this article in Table \ref{abbrev}. 

\begin{table}[h]
\centering
	\caption{List of Acronyms used in this survey.}
\begin{tabular}{p{6cm} p{2cm}}
		\hline
		\textbf{Terminology} & \textbf{Acronym} \\\hline
		
		Artificial Intelligence & AI \\
		Binarized Neural Network & BNN\\
		
		Capsule Network & CapsNet \\
		Conjunctive Normal Form & CNF \\
		Counter-Example Guided Abstraction Refinement & CEGAR\\
		Convolutional Neural Network & CNN \\
		Cyber-Physical System & CPS \\
		
		Deep Neural Network & DNN\\
		Denial-of-Service & DoS \\
		Generative Adversarial Network & GAN\\
		Internet-of-Things & IoT \\
		
		Linear Programming & LP \\
		Long Short-Term Memory & LSTM \\
		
		Machine Learning & ML \\
		Mixed Integer Linear Programming & MILP \\
		Multi-Layer Perceptron & MLP \\
		Neural Network & NN \\
		
		Recurrent Neural Network & RNN\\
		Satisfiability Modulo Theories & SMT\\
		Satisfiability solving & SAT solving\\
		Spiking Neural Network & SNN \\
		
		\hline
\end{tabular}
	\label{abbrev}
\end{table}

The rest of the article is organized as follows (see Fig. \ref{po}):
\begin{itemize}[leftmargin=*]
    \item Section~\ref{taxonomy} provides a taxonomy of ML that categorizes different network types, explains the ML design cycle, and introduces basic concepts about security and reliability.
    \item Section \ref{attacks} highlights the various attack strategies that jeopardize the integrity of ML systems, particularly at the cloud and edge levels, explains how these attacks are implemented, and identifies mechanisms that can lead to the mitigation of the attacks.
    \item Section \ref{defences} describes the defenses against the security attacks to secure ML systems, and states the shortcomings of these defenses.
    \item Section \ref{reliability} elaborates on the reliability concerns that reduce the accuracy of ML systems, specifically when deployed at the edge, in the absence of an explicit attacker.
    \item Section \ref{sec:mitigation} discusses the state-of-the-art techniques to mitigate the impact of the reliability issues in ML, and their corresponding limitations.
    \item Section \ref{fm} elaborates on the use of formal methods for Neural Network (NN) verification, presents the various types of verification techniques and their use for DNN verification in state-of-the-art, and explains the reasons for their limited success.
    \item Section \ref{challenges} discusses the open research challenges and perspectives towards designing secure and reliable ML systems. 
\end{itemize}

\section{Machine Learning: Concepts and Terminology} \label{taxonomy}

An ML system, like any other traditional system, takes in the input(s) and generates the corresponding output(s). However, unlike traditional systems, the ML system is capable of learning via input features and using the learned features in decision-making, which provide ML systems with the ability to perform tasks that are very challenging to perform using traditional systems.

NNs are often involved in the main decision-making of many modern ML systems. An NN comprises of an \textit{input layer} that connects the external environment to the ML system, an \textit{output layer} that outputs a decision, and \textit{hidden layer(s)} sandwiched between the input and output layers. State-of-the-art ML systems commonly use DNNs with two or more hidden layers. Each layer comprises of \textit{neurons/nodes}, which connect to other neurons in the corresponding layers via a non-linear \textit{activation} function. Each neuron has its associated parameters i.e., weight, bias, and/or filter coefficient. A detailed overview of neural networks can be found in \cite{sze2017efficient}\cite{lecun2015deep}.

\subsection{Neural Network Taxonomy}
If the input propagates through the network in only one direction, the network is said to be \textit{feed-forward}. If there are feedback loops in the network, the network is called a \textit{recurrent} neural network (RNN)\cite{rnn}. Long Short-Term Memories (LSTMs) \cite{lstm} are a branch of recurrent networks that retain information for a long duration, which makes them well-suited for time series prediction. When every neuron in one layer is connected to ``all'' neurons in the preceding layer, the network is said to be \textit{fully-connected}; otherwise, the network is \textit{sparse}. 

Since their advent, NNs have progressively improved over three generations (see Fig. \ref{tax}). The details of the NN types of each generation can be found in Appendix \ref{Appendx}.

\begin{figure*}[ht]
	\centering
	\includegraphics[width=\linewidth]{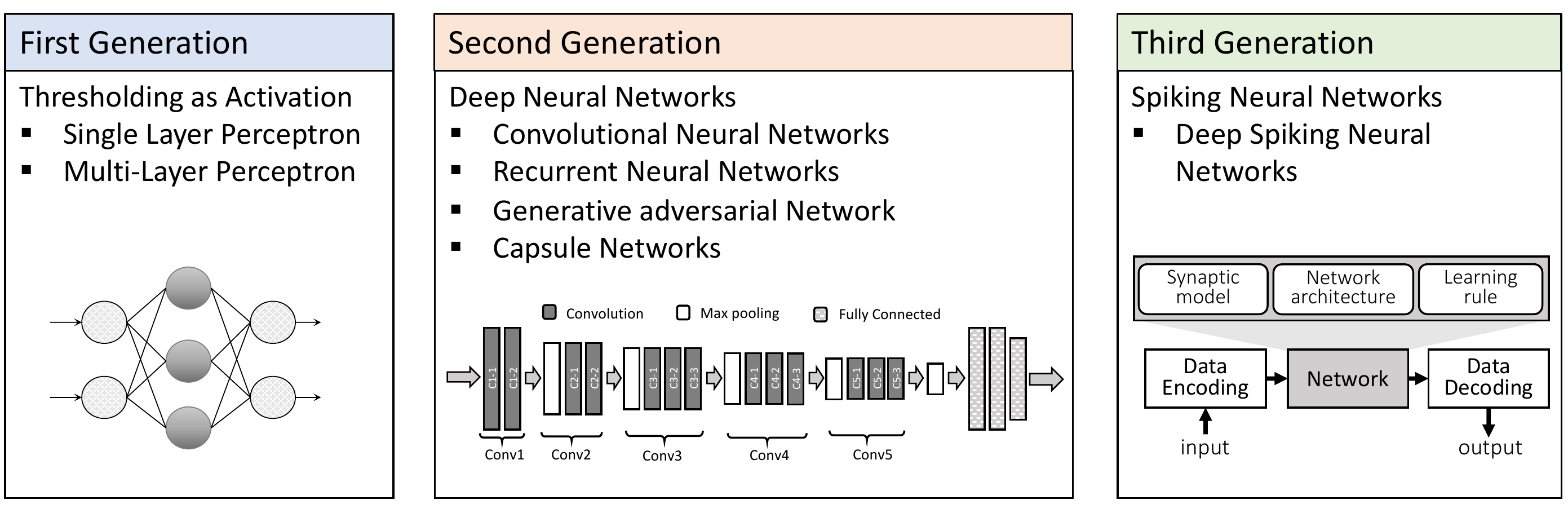}
	\caption{\textit{Summary of Neural Network models proposed over time.}}
	\label{tax}
\end{figure*}

\textbf{First Generation of NNs}:  The first generation of NNs~\cite{mcculloch1943} is comprised of single-layer and multi-layer perceptrons (MLPs)~\cite{mlp}. \textit{MLPs} are generally made up of multiple fully-connected layers connected with thresholding activations. 

\textbf{Second Generation of NNs}: To reduce the number of parameters in a network, this generation of NN introduces \textit{convolutional neural networks (CNN)}, which make use of convolutional layers comprising of convolutional filters to extract important features from the input, while providing a certain degree of shift invariance to the network \cite{lecun1990cnn}. A convolutional layer typically uses continuous non-linear activations, and it is often followed by a pooling layer. Pooling layers reduce the network parameters even further by retaining only the most important features from the preceding layer, which leads to information loss. 
This generation of NNs are increasingly being deployed in practical ML systems. 

A relatively new approach to solve the problem of information loss in CNNs is the use of \textit{capsule networks (CapsNets)} \cite{hinton2011capsnet}. 
CapsNets have hidden layers comprised of interconnected vectors that have input features and input probabilities, which allow these networks to learn spatial correlations between input features. As a result, CapsNets are able to infer high level entities quite similarly to human perception.

Another interesting approach towards NNs is the \textit{Generative Adversarial Networks (GANs)}~\cite{gans}. These networks make use of the simultaneous interplay between a generator and a discriminator, where the generator produces realistic synthetic inputs while the discriminator learns to differentiate between the real and synthetic inputs. This enables the NNs to generate synthetic outputs that are very difficult to distinguish from the real ones.

\textbf{Third Generation of NN}s: This generation of NNs makes use of \textit{Spiking Neural Networks (SNNs)} \cite{snn} in an attempt to emulate human brain like functioning. Unlike the networks discussed earlier, which consider the normalized firing frequency of neurons, SNNs use spike trains to mimic the spatio-temporal characteristics of the biological neurons.


\subsection{Neural Network Design Cycle}
\begin{figure}[t]
	\centering
	\includegraphics[width=\linewidth]{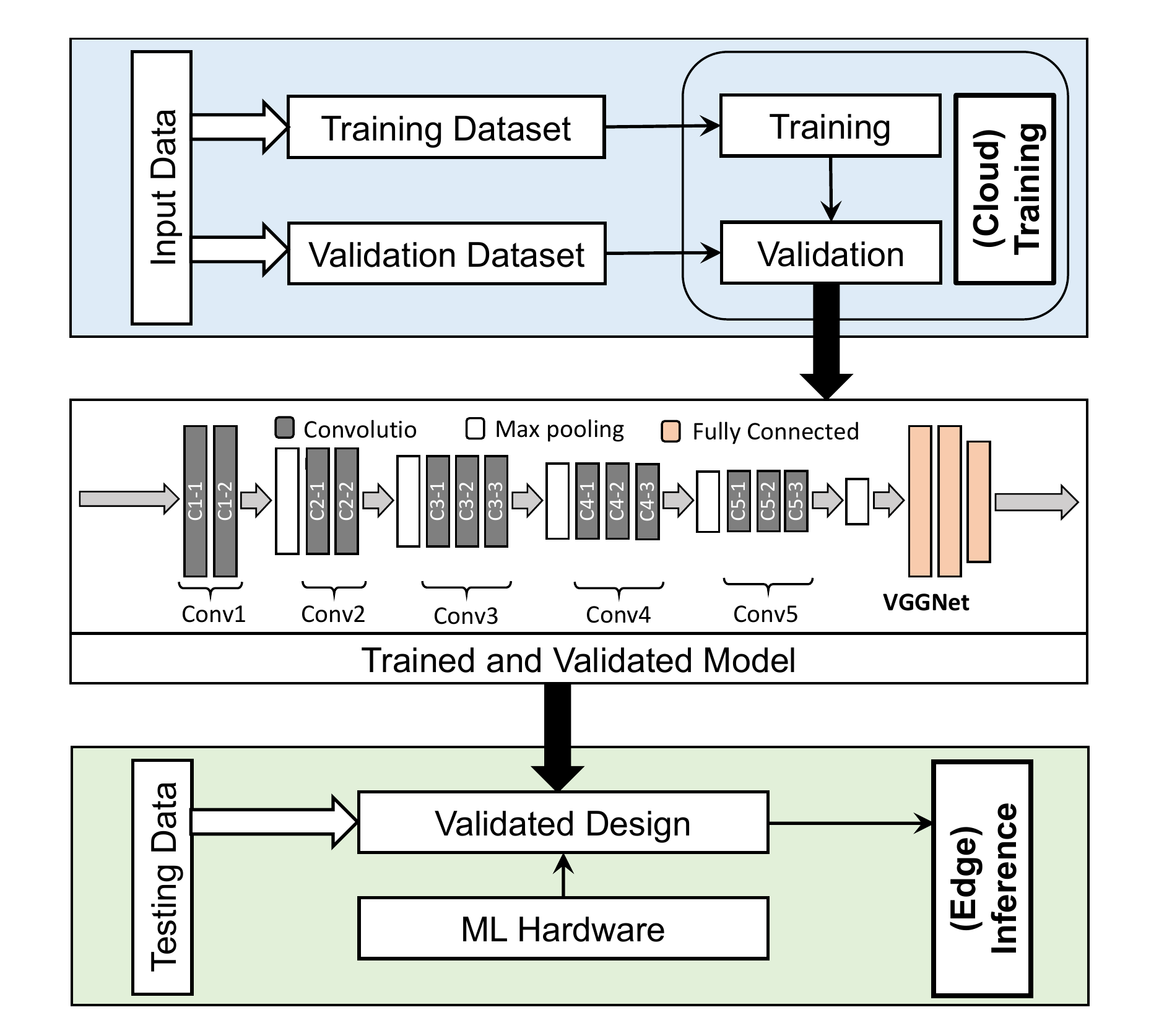}
	\caption{\textit{Design cycle of a NN-based ML system.}}
	\label{designcycle}
\end{figure}

Fig.\ref{designcycle} provides an overview of the NN-based ML design cycle, which can be categorized in the \textit{training} and  \textit{inference} stages. Training is typically performed at the Cloud, while inference is typically performed at the edge in real-world Smart CPS systems (e.g., autonomous vehicles and wearable healthcare devices). In certain IoT/CPS systems thT are not constrained with resources or real-timeliness, inference may also be performed at the Fog or Cloud (e.g., predictions on social networks and large-scale hospital data).

\textbf{Training:} Before deploying the NN into an ML system, the NN must be trained. \textit{Training} is a resource-intensive process, generally carried out by third party cloud servers, which involves the use of a \textit{training dataset} to find suitable values for the network parameters. Training is composed of a forward-pass and a backward-pass. The forward-pass calculates the predicted output values by propagating inputs through the network, using the current parameter values. The backward-pass updates the network parameters while minimizing the loss function associated with correct and predicted output values. This process (i.e., a forward-pass and a backward-pass), when repeated once for all the samples in the training dataset, is called an \textit{epoch}. The overall training process of a NN involves several \textit{epochs}. 

At the end of each \textit{epoch}, the accuracy of the network is analyzed for some unseen data, which is not part of the training dataset, i.e., the validation dataset. The result of this testing can be used to fine tune the network hyper-parameters, like the number of layers, and select the best trained model. The training process then resumes, and the network parameters are again updated using the training dataset until either the process reaches the maximum number of epochs (or cycles), or the network reaches the desired level of accuracy with the validation dataset.

The most common way to check the final inference accuracy of a trained network is to use a \textit{testing dataset}.
If the trained network is able to classify testing inputs correctly for more than the desired number of testing inputs, the network is considered suitable for deployment into a practical system. However, a DNN might misclassify an input that is perceptually similar to another input correctly identified by the same DNN~\cite{AdvAttack}. To ensure the security, reliability and safety of ML systems for safety-critical applications, e.g., autonomous vehicles and smart healthcare, it is imperative to develop a framework to analyze and verify these critical misclassifications. An orthogonal research direction, therefore, is to use formal verification for ascertaining the dependability of the trained DNN.

\begin{figure}[t]
	\centering
	\includegraphics[width=\linewidth]{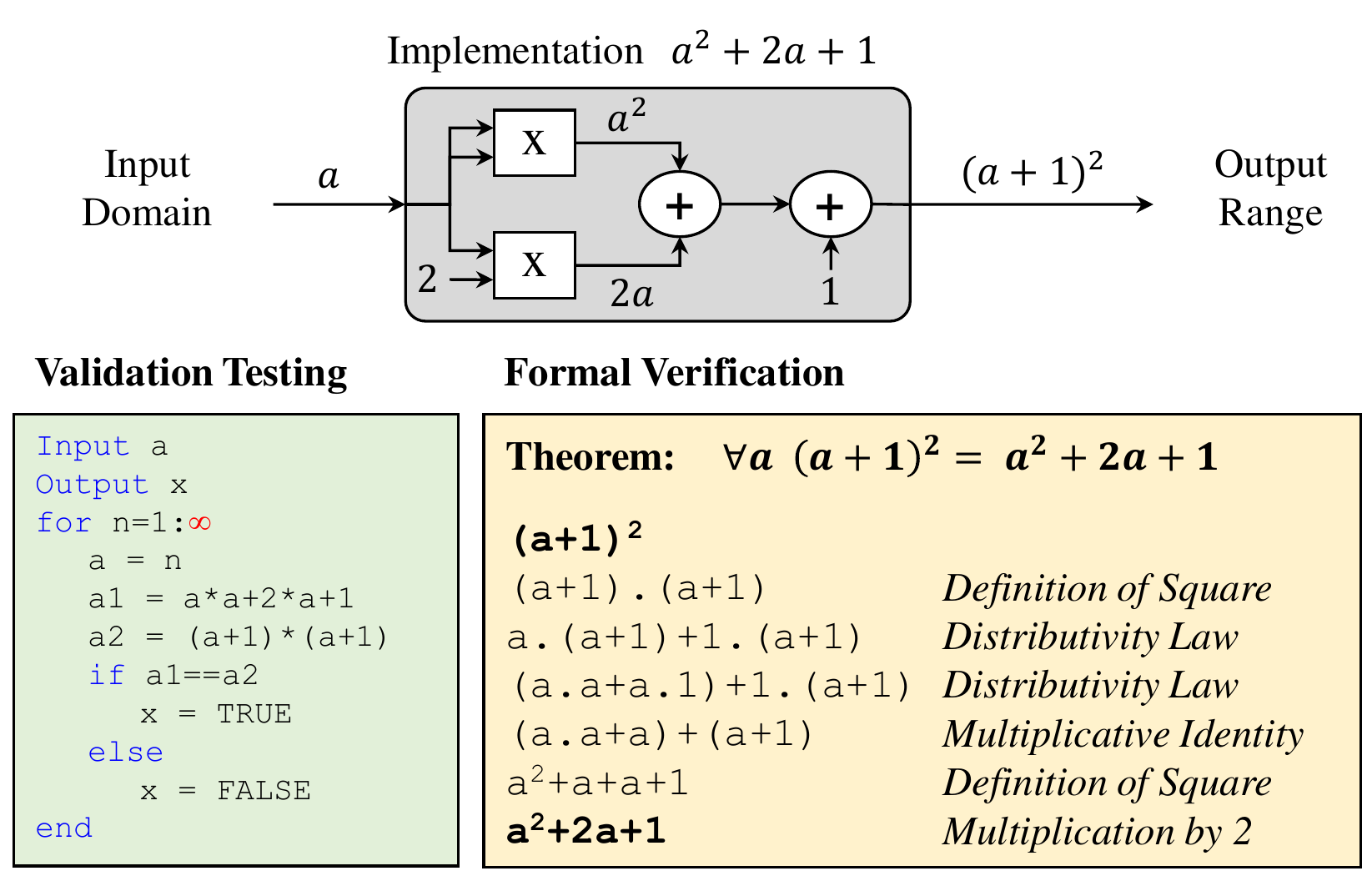}
	\caption{\textit{Comparison between testing and verification for a small hypothetical system: ensuring behavioral correctness of the system for all possible inputs is not always feasible with testing.} }
	\label{verification1}
\end{figure}

\begin{figure}[ht]
	\centering
	\includegraphics[width=\linewidth]{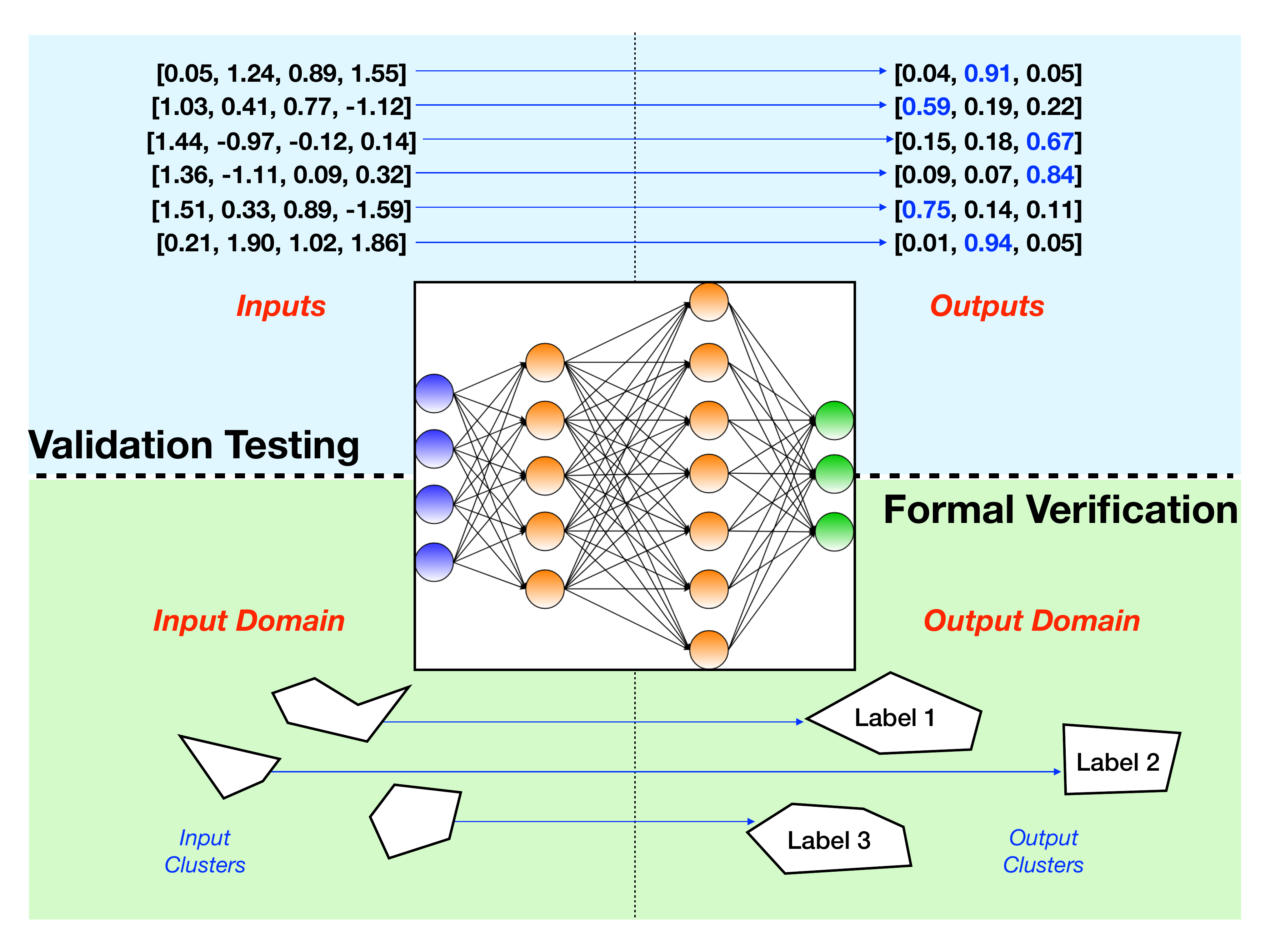}
	\caption{\textit{Comparison between testing and verification for a neural network based system: Verification is intended to determine whether the bounded inputs are reachable to the correct output bounds.}}
	\label{verification2}
\end{figure}

Although an established research domain \cite{FMforHW}\cite{FMforSW}, \textit{formal verification} started gaining interest in the ML research community only since the last decade. Formal verification is an approach to check the correct behavior of a system on the basis of sound mathematical reasoning. Unlike testing, verification provides guarantees regarding system accuracy, independent of ``specific" system inputs. Hence, as shown in Fig.~\ref{verification1}, the guarantees provided by verification are valid for the entire (infinite) input domain whereas those provided by testing are limited only to the (finite) tested data. In terms of ML systems, due to the complexity of the underlying system, the objective of verification is usually to verify the correctness of the network for bounded input regions, as demonstrated in Fig. \ref{verification2}, rather than for the entire input domain. 

\textbf{Inference:} A trained and tested/verified NN can be deployed in a real-world ML system. At this stage, the NN performs classification/decision-making using actual, previously unseen, data (i.e., in real-time). ML inference is carried out at the edge of the IoT/CPS system, hence exposing the system to numerous security and reliability concerns during the operations under varying scenarios and harsh environmental conditions.

\subsection{Robustness} 
A common term associated with the performance of DNNs is robustness. \textit{Robustness} is the DNN property that determines the integrity of the network under varying operating conditions, and the accuracy of DNN outputs in the presence/absence of input or network alterations. This can be divided into two sub-properties: security and reliability \cite{kriebel_survey}. The DNN is said to be \textit{secure} against an attack if the attacker cannot steal information (via IP stealing or side channel attack), engage the system resources (e.g., using hardware intrusion or Denial-of-Service (DoS) attack), modify the network parameters (e.g., by inserting hardware or neural-level Trojans), or render an incorrect input to the DNN (e.g., using an adversarial attack). In case of reliability, there is no explicit attacker. The network is said to be \textit{reliable} if it does not display any changes to its output, parameters, or behavior, due to the changes in environmental conditions, during fabrication and deployment.

\section{Security Vulnerabilities of ML systems} \label{attacks}
As hinted in the previous section, despite being highly sophisticated in learning and decision making, ML systems are very vulnerable to attacks. Depending on the type and intensity of the attack(s), and the application where the system is deployed, these ML vulnerabilities can lead to slight discrepancies in the result, or can lead to lethal consequences in a safety-critical application \cite{auto-drive1}. This section describes the most common security issues in ML systems and DNNs at the cloud and the edge, as summarized in Table~\ref{summary-sec}).

\begin{table*}[t]
	\centering
	\caption{\textit{Summary of the various security threats and their countermeasures for ML-based systems.}}
	\label{summary-sec}
	\begin{tabular}{|m{2.4cm} | m{1cm} | m{1cm} | m{1.7cm} | m{3.3cm} | m{6.25cm} |}
		\hline
		
		\multirow{3}{*}{\textbf{Threat}} & \multicolumn{3}{|c|}{\textbf{Insertion Point}} & \multirow{3}{*}{\textbf{Vulnerability}} & \multirow{3}{*}{\textbf{Countermeasures}} \\ 
		\cline{2-4}
		& \multicolumn{2}{|c|}{Design Phase} & \multirow{3}{*}{DNN Inference} & &\\
		\cline{2-3}
		& \textit{Hardware Design} & \textit{DNN Training} & & & \\ \hline
		
		Adversarial Attack & & & \ding{51} & Input & Gradient Masking \cite{GradientMasking}, Pre-Processing Filters \cite{fadeML}, ~Adversarial Retraining\\ \hline
        
        Backdoor Attack & & \ding{51} & & Network Parameters (W, b) & Pruning \cite{badnet}, Fine Tuning \cite{Prune-aware-attack}\\ \hline

        Data Poisoning & & \ding{51} & \ding{51} & Input & Encryption \cite{Encryption1}\cite{Encryption2}\cite{Encryption3}, Local Training\\ \hline
   
        IP Stealing & & & \ding{51} & System Response & Obfuscation, Encryption \cite{Encryption-ML-confidential}\\ \hline
        
        Hardware Trojan & \ding{51} & & & Hardware, System Response & Equivalence Checking \cite{book_SWforHW}, Side-Channel analysis \cite{SideChannel}\\ \hline

        Side-channel Attack & & & \ding{51} & System Response & Randomness \cite{SideChannelCNN}\cite{sidechannel-modelstealing} \\ \hline
        
        \end{tabular}
\end{table*}

\subsection{Adversarial Attack}
\begin{figure}[ht]
	\centering
	\includegraphics[width=\linewidth]{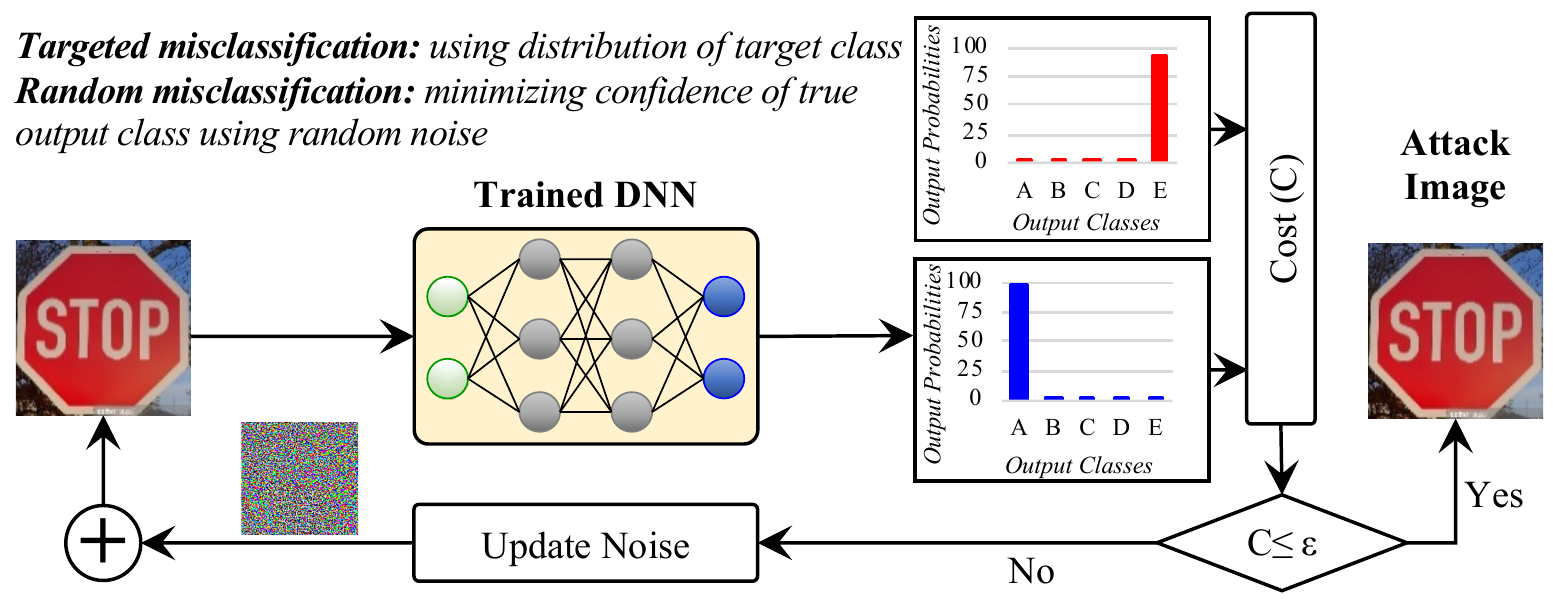}
	\caption{\textit{Adversarial Attack on a trained DNN: an adversarial attack can result in the misclassification (either targeted or random) of traffic sign boards, which is a concern in autonomous driving \cite{trisec}}.}
	\label{adv}
\end{figure}

Since their discovery, adversarial attacks~\cite{AdvAttack} have been a widely studied DNN security threat~\cite{FGSM}\cite{capsnet-advattack1}\cite{capsnet-advattack2}. In an adversarial attack, the known DNN parameters are exploited to minimize the cost function corresponding to noise patterns $\delta x$, which, when added to the input $x$, can cause \textit{misclassification}, as shown in Fig.~\ref{adv}. The noise added is usually imperceptible, making the task of distinguishing between clean and malignant inputs nearly impossible. This can be represented formally as:
\begin{equation}
    f(x) \neq f(x + \delta x +EN) ~~s.t.  ~~\delta x \leq \epsilon
\end{equation}
where $EN$ represents the noise existing in the physical environment even in the absence of an explicit attacker.
The adversarial noise can lead to either a random incorrect output class, i.e., an \textit{untargeted attack scenario}, a specific calculated output class, i.e., a \textit{targeted attack}, or simply reduce the confidence of the correct output class~\cite{noiseML}, i.e., \textit{confidence reduction}.

Adversarial attacks can be categorized as either \textit{evasive} or \textit{poisoning}~\cite{adv-categorization}, depending on the access of the attacker to the DNN design cycle. In \textbf{evasive attacks}, the attacker has no access to the DNN training process and training dataset. The attack is solely configured during the DNN inference stage, using either input gradients, output probability vectors, or the output decision \cite{boundary-attack,brendel2017decision,chen2019boundary,cheng2018query,pengcheng2018query,dong2019efficient,khalid2019red}. For instance, the Fast Sign Gradient Method (FSGM)~\cite{FGSM}, determines the direction of the loss function via the input gradient, scales down its value, and adds the noise to the input. In the Jacobian Saliency Map Approach (JSMA)~\cite{JSMA}, the input gradient (Jacobian) is again used, but the objective is to add the noise to a subset of input nodes, sufficient for misclassification. Other works \cite{IFGSM}\cite{CW} make use of input gradients to propose adversarial attacks. TrISec \cite{trisec} improves the imperceptibility of an adversarial attack by introducing a new methodology that uses additional parameters (e.g., correlation coefficient between the target image and the original image,  and structural similarity index) in the DNN training algorithm.
Works like \cite{red}\cite{boundary-attack} make use of output labels to determine attacks in close proximity to the classification boundary.

\begin{figure}[ht]
	\centering
	\includegraphics[width=\linewidth]{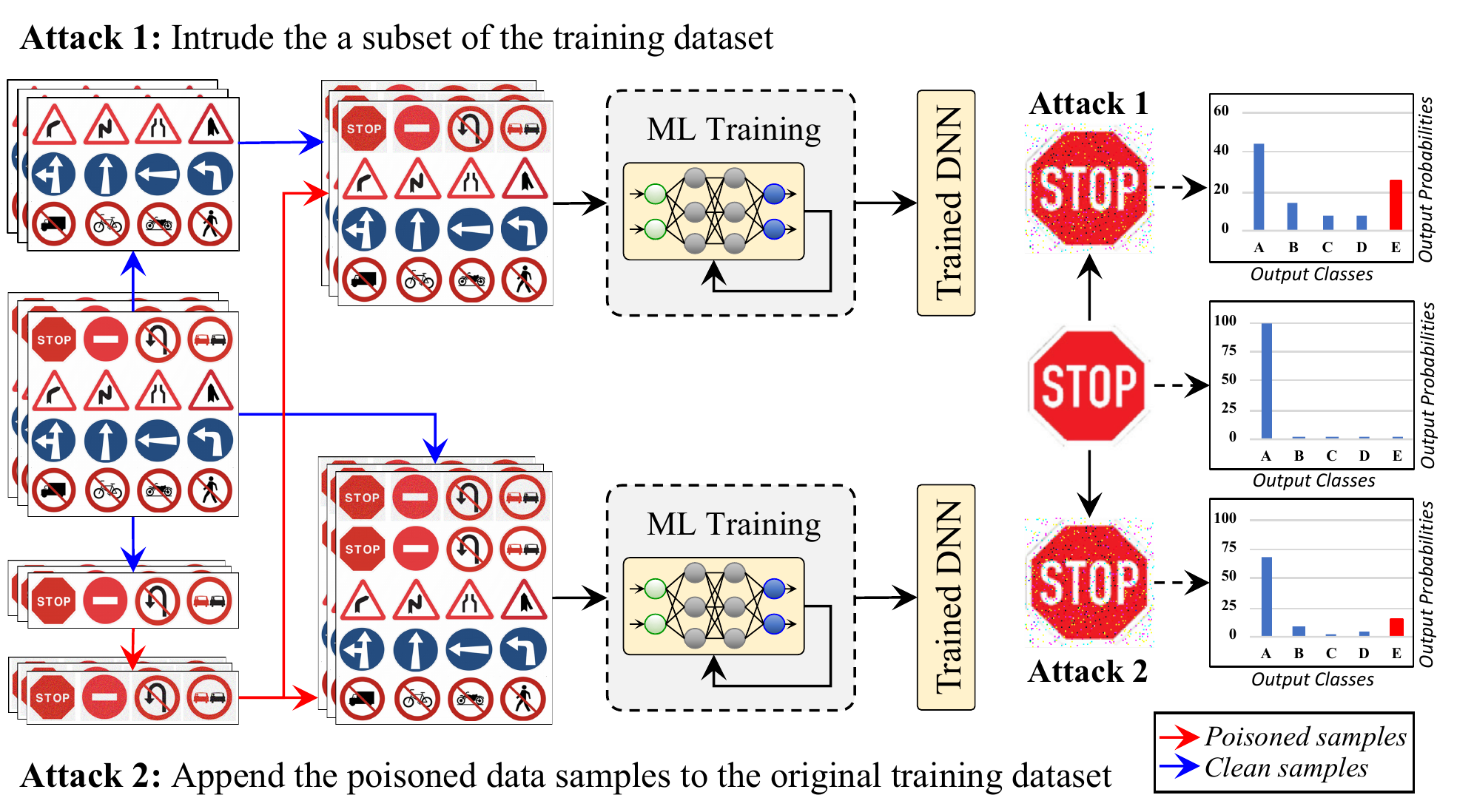}
	\caption{\textit{Classification accuracy of DNN trained on a poisoned dataset.}}
	\label{poisoning}
\end{figure}

\begin{figure}[ht]
	\centering
	\includegraphics[width=\linewidth]{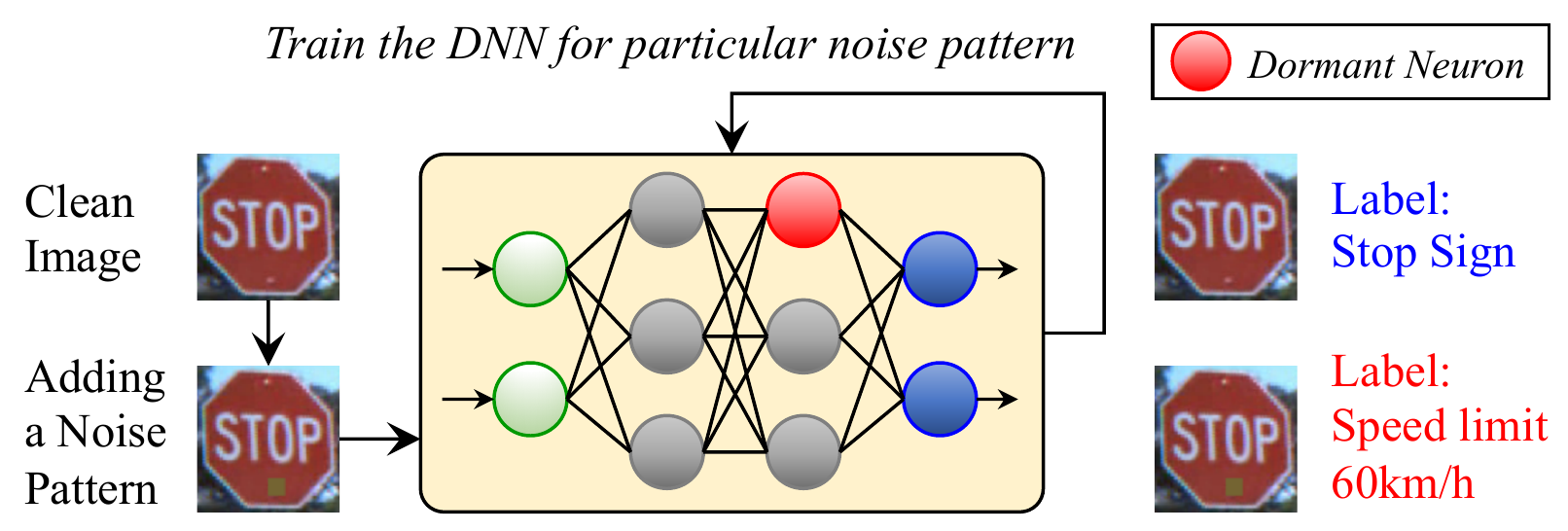}
	\caption{\textit{Effect of a backdoor on DNN accuracy. The dormant neurons (red) learn to associate the backdoor with a targeted misclassification label.}}
	\label{backdoor}
\end{figure}

In \textbf{poisoning attacks} \cite{DataPoisoning}, the attacker has access to the training dataset/training procedure. The attack is implanted in the DNN during training by feeding the network with malicious training data. Fig.~\ref{poisoning} shows two example poisoning attacks that increase the probability of misclassification of an stop signal (red bars). The data could be poisoned with tailored noise~\cite{badnet}\cite{neural-trojan}, also known as \textit{backdoor attack}, or simply through random noise~\cite{noiseML}. Sparsity of the network accounts for the success of poisoning adversarial attacks. Dormant  neurons  in  a  trained  DNN  have  weights  and  biases too  small to  be  of  any practical  significance to the output  calculation. The existence of such neurons signify that the network has the capacity to learn more. Hence, such networks can be trained on poisoned data (as shown in Fig.~\ref{backdoor}). The DNN behaves correctly for the clean data but exhibits a malignant behavior for the poisoned data.  A recent work demonstrates the use of poisoning (with noisy image patches) to either misclassify humans as different objects or completely hide a person from the object detection system \cite{YOLO:attack:CVPR2019}.

For most of the adversarial attacks, a common inadequacy is to ignore the pre-processing filtering stage in an ML system \cite{fadeML}. The pre-processing stage generally employs different averaging filters, to smooth out any noise in input. This undermines, if not completely eliminates, the threat of misclassification via adversarial attacks.

\subsection{Neural-Level Trojans}
Another class of attacks, the neural-level trojans \cite{potrojan}, involves the insertion of additional neurons into a pre-trained DNN by third-party training servers. The number of extra neurons must be minimized to avoid raising suspicion regarding the DNN model. Conceptually, similar to hardware trojans in system hardware (discussed later) and backdoor attacks, the additional neurons in neural-level trojans trigger malicious DNN behavior only when prompted by specific inputs. However, most of these attacks require to re-train the network and use complex internal triggering mechanisms.

\subsection{Hardware Attacks}
\textbf{Hardware trojans} \cite{HWtrojan1}\cite{HWtrojan2}\cite{trojanDoS}\cite{NN-hw-trojan}\cite{trojanIPsteal} are malicious components implanted into the system hardware, which compromise the security of a ML system. Hardware trojans can introduce undesired system behavior, or be dormant in the normal system operation and be triggered at a specific instance. They may leak system information, thus aiding IP stealing (discussed later), or simply consume system power and resources.

The attack is usually instigated by an untrusted manufacturer/foundry, at the manufacturing stage of the system lifecycle. The size of the trojan is usually small, and hence goes unnoticed. Often, the overall number of components on the chip is kept unchanged and the power trace of the trojan is also minimized \cite{trojanzero}, to ensure a successful stealthy attack.

\begin{figure}[ht]
	\centering
	\includegraphics[width=\linewidth]{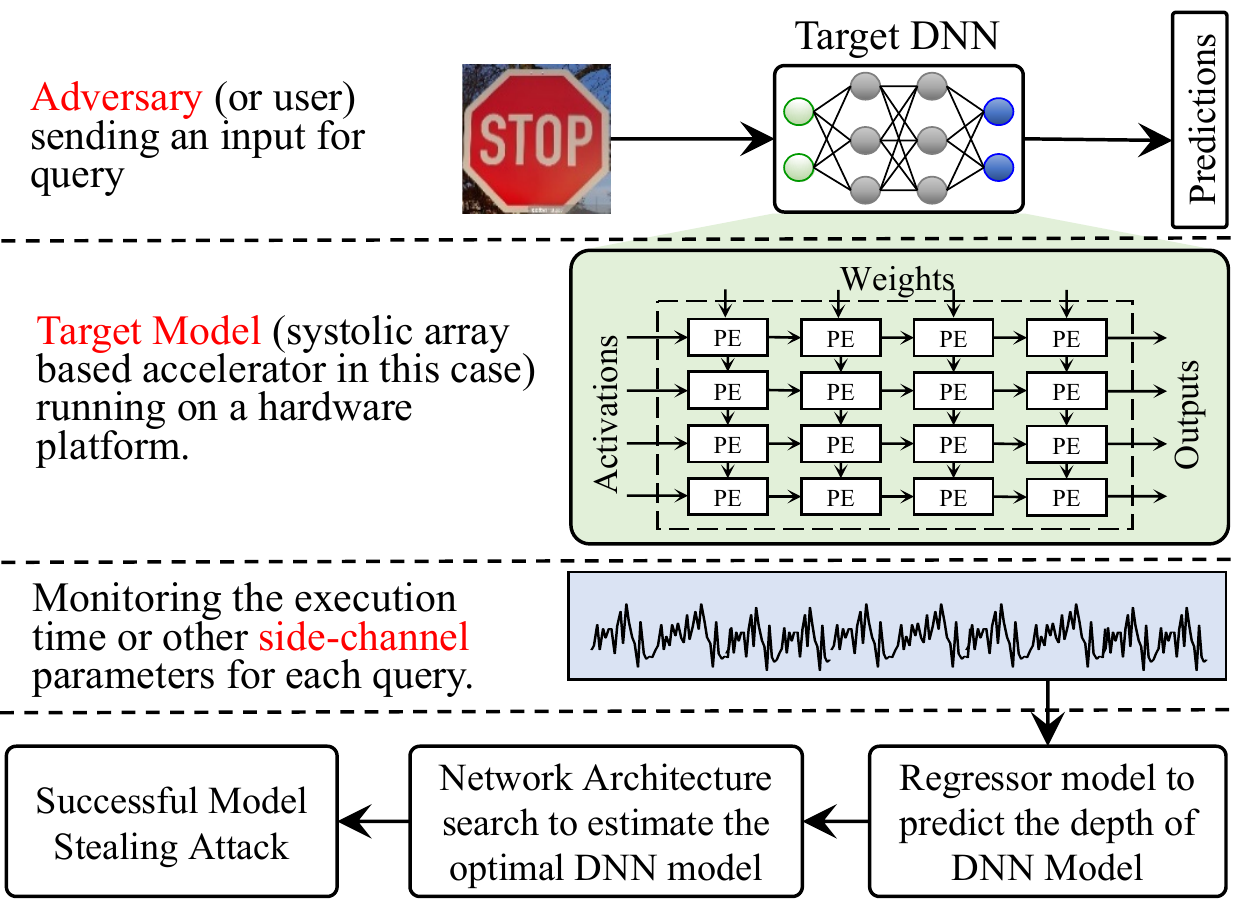}
	\caption{\textit{Side-channel attack based on the execution time of individual input queries, which can be used to decipher the depth of the DNN model and estimate the network parameters/model.}}
	\label{sidechan}
\end{figure}

\textbf{Side channel attacks}, as shown in Fig.~\ref{sidechan}, are another type of hardware attack that is crafted using leaking information from the system hardware. Most systems leak information via side channels such as components' power consumption~\cite{SideChannelCNN,sidechannel-modelstealing,SideChannel,duddu2018stealing,batina2018csi,yoshida2019model,pal2019framework}. This information can be analyzed and used to 1) compromise the security and privacy of the system, and 2) reverse engineering and steal the model parameters~\cite{batina2019csi,hua2018reverse}.

Analyzing the different side-channels of a system enables to target different parameters of an ML system. 
For instance, the leaking power traces close to the input of the DNN provide clues regarding system input, whereas the information regarding execution times provide predictions for the network architecture \cite{SideChannelCNN}\cite{sidechannel-modelstealing}. However, a common limitation with most side-channel attacks is assuming the absence of noise in the system. Inclusion of noise in the side-channel attack's threat model generates randomness in the leaked information,  which reduces the chances of a successful attack.

\subsection{IP Stealing}
\begin{figure}[ht]
	\centering
	\includegraphics[width=\linewidth]{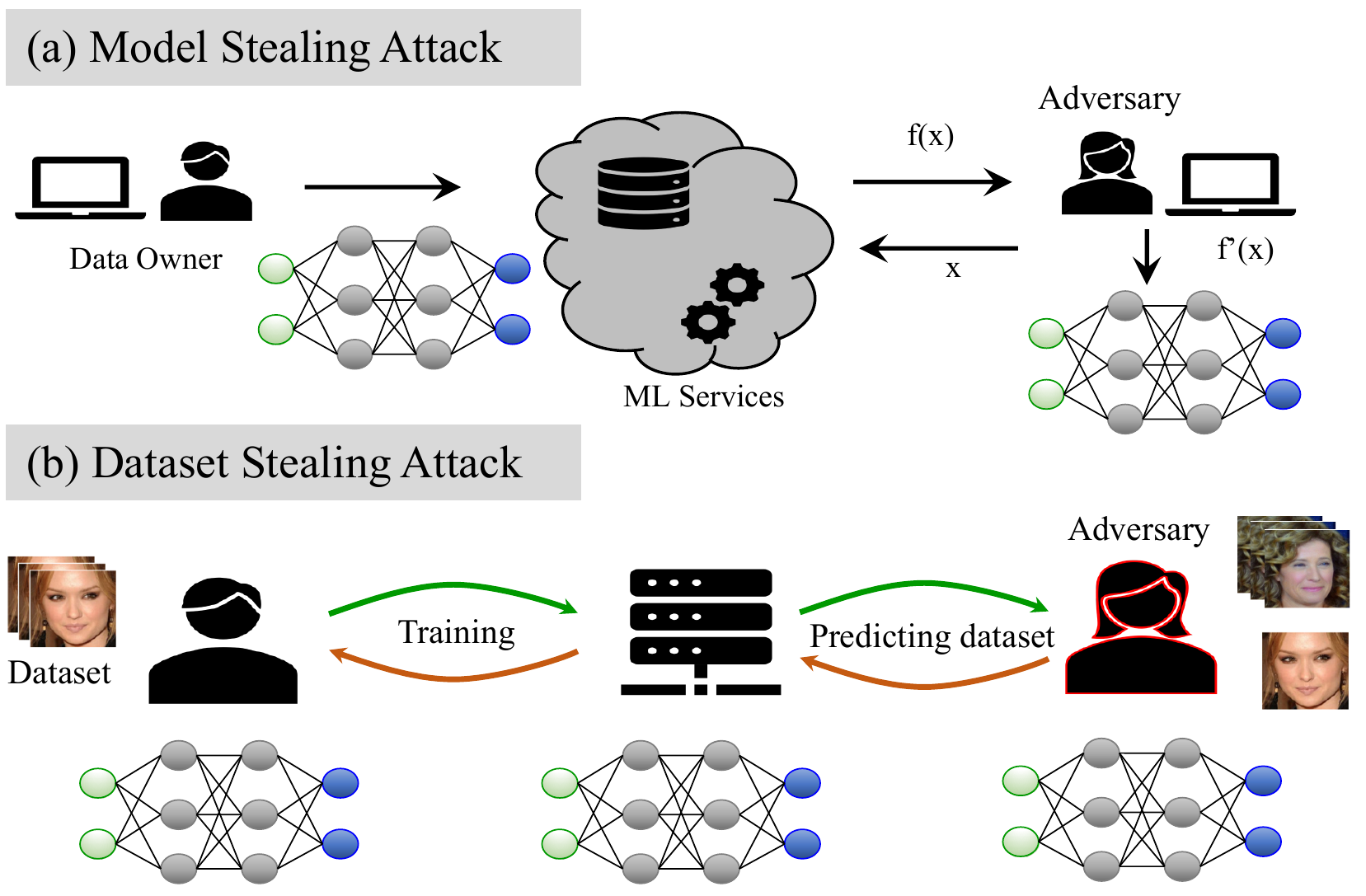}
	\caption{\textit{IP Stealing from a trained DNN: the objective of a stealing attack can either be: (a) to estimate the underlying DNN model, or (b) predict the dataset used for DNN training, using multiple queries.}}
	\label{steal}
\end{figure}

Attacks to steal Intellectual  Property  (IP)  are  another significant security threat for ML systems. IP stealing involves determining either the underlying \textit{model} of the ML system (\textbf{Model Stealing Attack}), possibly without any access to the description or internal parameters of the system \cite{Model-Steal}, or predicting the \textit{data} the DNN was trained on using the available model description (\textbf{Dataset Stealing Attack})~\cite{Model-Inversion}. Both types of attacks are shown in Fig.~\ref{steal}. Leaking side-channels of the model, responses of queries to the system, and similar  behavioral network characteristics can be exploited, analyzed, and reverse engineered to obtain the underlying IP.

\section{Defenses Against Security Vulnerabilities of ML Systems} 
\label{defences}
To ensure correct operation in the presence of security attacks, several security defenses have been proposed over the years. This section describes some of the most prominent ML system defenses against security threats, categorized according to the threats they counter.

\subsection{Defending Against Adversarial Attacks} \label{adv-defenses}
\begin{figure}[ht]
	\centering
	\includegraphics[width=\linewidth]{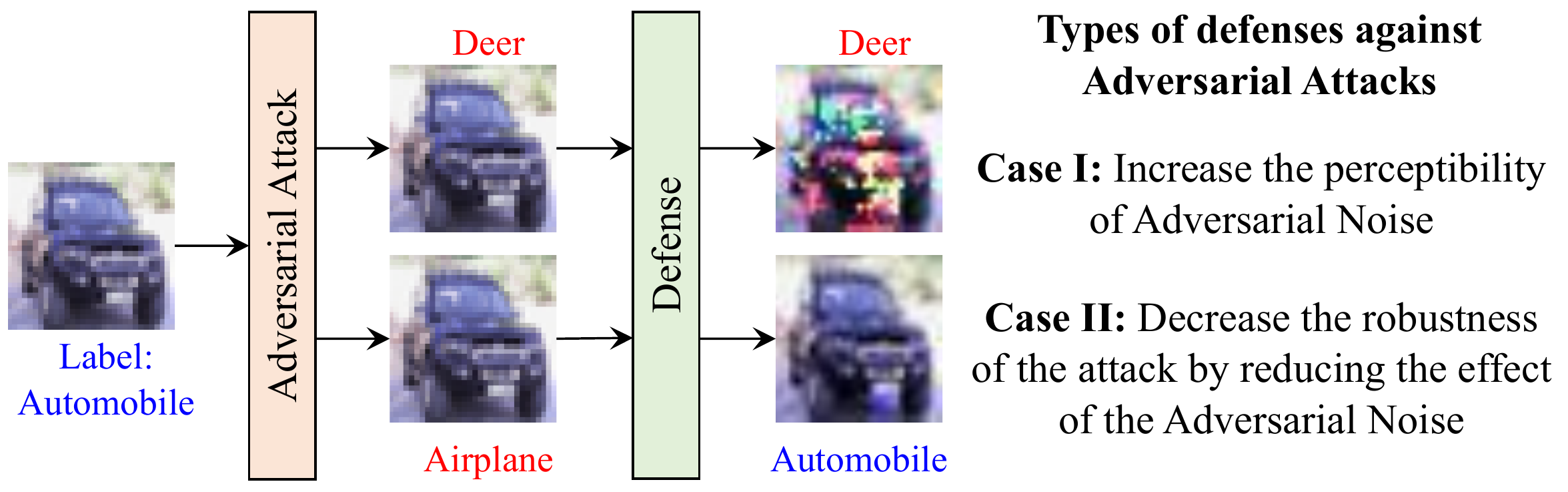}
	\caption{\textit{Defenses against adversarial attacks either increase the  perceptibility of adversarial noise (Case I) or decrease the effect of the adversarial noise (Case II).}}
	\label{def-obj}
\end{figure}

The concerns originating from adversarial attacks are confidence reduction of the true output class and misclassification. As shown in Fig. \ref{def-obj}, the defenses against adversarial attacks are generally intended either to 1) increase the perceptibility of the attack, thereby ensuring that the clean and malignant inputs are perceptually distinguishable, or 2) reduce the impact of the attack by enhancing the DNN's robustness against it.

\begin{figure}[ht]
	\centering
	\includegraphics[width=\linewidth]{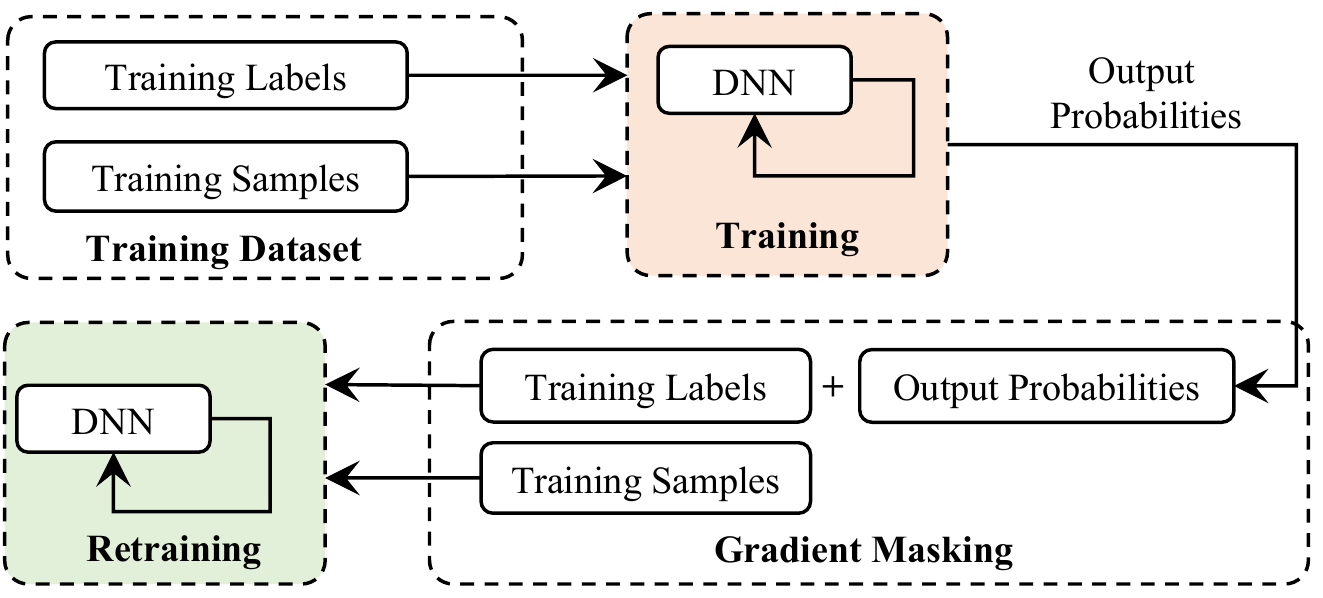}
	\caption{\textit{Using Gradient Masking to hide the input gradients that might be used by the attacker to determine the perturbations that need to be inserted to perform the Adversarial Attack.}}
	\label{mask}
\end{figure}

\begin{figure}[ht]
	\centering
	\includegraphics[width=\linewidth]{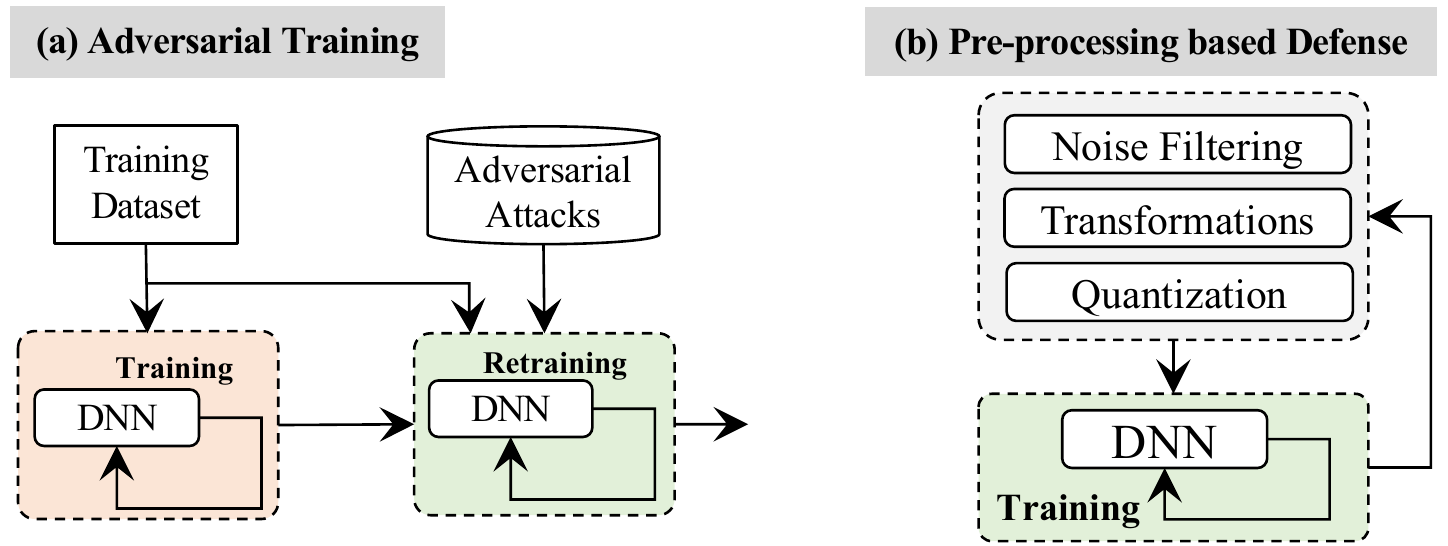}
	\caption{\textit{(a) Improving a DNN's accuracy in the presence of a known attack by training the dataset with Adversarial Examples obtained from known adversarial attacks, i.e., Adversarial Training. (b) Reducing the effects of adversarial noise added to the input via input pre-processing techniques such as noise filtering, quantization and other input transformations.}}
	\label{ATraining}
\end{figure}

For evasion-based adversarial attacks crafted using input gradients, a natural defense strategy is to hide these gradients using a technique called \textbf{gradient masking} \cite{GradientMasking}. This technique, as explained in Fig.~\ref{mask}, reduces the dependability of output classification by retraining the DNN with the output probability vector. \textbf{Adversarial training}, as shown in Fig.~\ref{ATraining}(a), is another commonly used defense \cite{deepxplore}\cite{retraining}, where a trained DNN is retrained with adversarial inputs and the correct corresponding output labels. This improves the accuracy of the system in the presence of a \textit{known} attack. Another defense, which actually constitutes a part of most practical ML systems, is the use of \textbf{input pre-processing} \cite{fadeML}. This defense smooths out, transforms and truncates the noise before it is even fed to the DNN. As shown in Fig.~\ref{ATraining}(b), this defense reduces the adversarial noise and hence reduces the chances of a successful attack. A recent defense against adversarial attacks is to train \textbf{robust image classifiers} \cite{DropPix:CVPR2019}. 
This defense exploits the fact that images contain high redundancy due to the strong correlation between neighboring pixels, so that a subset of pixels can be used to represent the same information. This subset is chosen by randomly dropping pixels from input images, and it is used during DNN training and inference. The drop rates are chosen randomly between 0\% and 100\% for each input image and at each epoch. The model trained on such subsampled datasets is robust against adversarial attacks. 


Most of the above defenses may work against a naive attack. However, for a strong attack, the defenses may fail. Many studies show that gradient masking does not increase the robustness of a DNN~\cite{clever+,goodfellow2018gradient,athalye2018obfuscated}, and hence can be broken with the use of a substitute model to identify the approximate gradient direction \cite{substitute-model}. Pre-processing aware attacks \cite{fadeML} can break the filtering defense. Likewise, as studied by several works \cite{retraining_fails}\cite{bastani}, adversarial training overfits a DNN to the adversarial examples and does not necessarily make the network more robust. Hence, a stronger attack can again make the DNN fail for certain inputs.

For poisoning-based adversarial attacks, a simple defense strategy is not to outsource the training process to a third party  (i.e., \textbf{local training}). 
However, training is a lengthy process, requiring large computational resources. Hence, local training is not always feasible for large DNNs. To outsource the training of large DNNs, the training data can be \textbf{encrypted} before outsourcing it to the third party \cite{Encryption1,Encryption2,Encryption3}, to overcome the impact of data poisoning. 

For attacks exploiting dormant neurons in the network, \textbf{pruning} can be employed to remove the (dormant) neurons that are not significant to the network inputs, reducing the chances for a successful backdoor attack. Yet, pruning-aware attacks \cite{Prune-aware-attack} can be used to train only the significant network neurons with  backdoor behavior, which eliminates the effectiveness of the pruning defense. Another defense is to \textbf{fine tune} the DNN with clean inputs~\cite{Prune-aware-attack}. Although this does not eliminate the backdoors from the network, it significantly reduces the chances of a successful backdoor attack.

To formulate better defenses against adversarial attacks, a current research focus is to determine robustness bounds for DNNs using formal methods \cite{BNNverification,dutta2018,MIT}. Although this area of study is relatively new and generally not scalable to practical DNNs, it has the potential to determine the actual boundaries where the DNNs will not be vulnerable to the adversarial attacks anymore. However, the question of how the knowledge of these bounds can be used to actually prevent adversarial attacks is yet to be answered.

\subsection{Defending Against Neural-Level Trojans}
Similar to adversarial attacks, the trigger for incorrect DNN behavior in Neural-Level Trojans is a malicious input. Hence, techniques that manipulate or detect input discrepancies can reduce the effect of neural-level trojans. Such approaches include  \textbf{input pre-processing} \cite{fadeML} to smooth out the input trigger, input \textbf{anomaly detection} \cite{anomaly-detection} to identify suspicious input patterns, and \textbf{prediction distribution} \cite{neural-trojan} to identify the bias of DNN towards the targeted output. Since trojans are inserted into pre-trained DNN models, their effect could also be negated using \textbf{local training} \cite{Prune-aware-attack}, i.e., training the DNN model locally instead of outsourcing the training process to third party cloud servers.

\subsection{Defending Against Hardware Attacks}
\begin{figure}[ht]
	\centering
	\includegraphics[width=\linewidth]{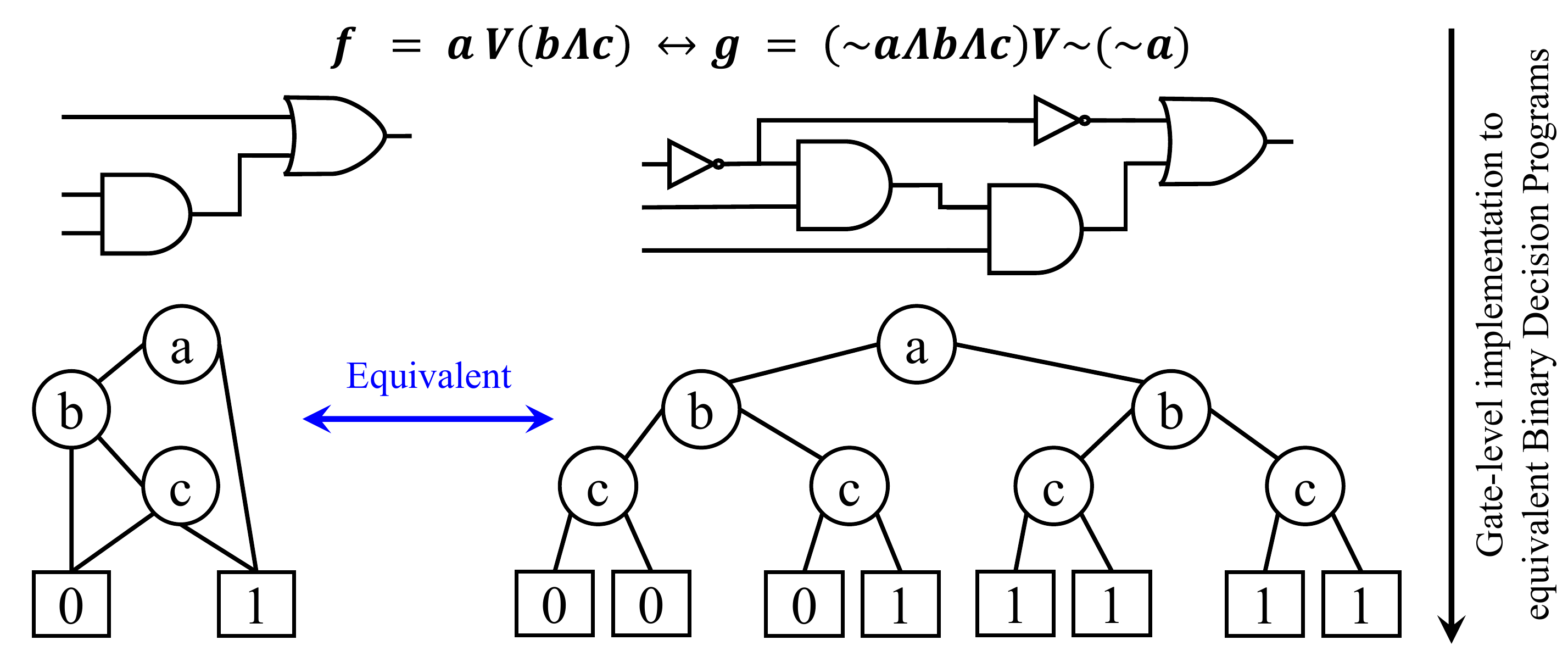}
	\caption{\textit{Using Binary Decision Diagrams (BDDs) for hardware equivalence checking.}}
	\label{equiv}
\end{figure}

Hardware trojans~\cite{agrawal2007trojan,adee2008hunt,HWtrojan1,jin2008hardware,tehranipoor2010survey} are a hardware-related security problem in ML systems. A hardware trojan is  a malicious modification of a circuit design that results in an undesired behavior, e.g., leakage of sensitive information, malfunction, or performance degradation. Since these attacks make use of hardware modifications, a suitable defense strategy against them is to use \textbf{formal methods}~\cite{FMforHW}, particularly via equivalence checking. Fig.~\ref{equiv} demonstrates the use of Binary Decision Diagrams (BDDs) for equivalence checking \cite{book_BDD} of simple gate-level circuits. The biggest obstacle to implement the equivalence checking defense is the absence of a golden/reference model of the actual system hardware, to compare with the intended system model \cite{HWtrojan1}\cite{HWtrojan2}.

Other potential defenses against hardware trojans include \textbf{side-channel analysis} \cite{SideChannel} for anomaly detection, and cross-layer attack modeling via bridging the gap between the hardware and software~\cite{book_SWforHW}. This defense often assumes that the leaking information, e.g., power trace, of the trojan is large enough to be detectable. The defense becomes ineffective when the assumption does not hold~\cite{trojanzero}.

As mentioned in the previous section, side-channel attacks make use of side-channel leakage from the system, often giving rise to other security vulnerabilities in ML systems, such as hardware intrusion \cite{noiseML} and IP stealing \cite{SideChannel}. Side-channel attacks often rely on the exactness of the leaking information, hence, the defense against them rely on the addition of random \textbf{noise} to system operations. For instance, a random selection of the next operation, whenever the sequence of operations does not matter, like selecting the sequence in which the image pixels are fed to the adder in a NN, could potentially make the inference of useful knowledge from side-channels more difficult \cite{MaskedNET}\cite{SideChannelCNN}\cite{sidechannel-modelstealing}. 

\subsection{Defending Against IP Stealing}
The most common IP stealing attacks involve stealing private or secret information (privacy infringement) and the robbery of the IP (piracy).

To protect \emph{privacy} of data, the simplest defenses include \textbf{blurring}, \textbf{obfuscation}, and even the addition of \textbf{adversarial noise} to the data \cite{obfuscation}\cite{FaceID}. In practice, these approaches may not work well as they may not be strong enough \cite{StealingLimitation}. Relatively stronger defenses include the use of \textbf{encryption} \cite{Encryption1}\cite{Encryption2}, i.e., data confidentiality, while outsourcing the data for training. Similarly, measures to ensure IP privacy during third-party DNN training include the use of multiple training servers for joint dataset \cite{safetynets}, verifying the training procedure \cite{secureml}, ensuring privacy after training by network transformation \cite{Minionn}, obfuscating defenses against reverse engineering-based attacks \cite{isakov2018preventing,liumitigating}, and isolating the hardware accelerators \cite{wang2019npufort}.

To protect IP against \emph{piracy}, the \textbf{rounding} approach \cite{ModelStealing} can be a potential defense. The leaking side channels could be a potential vulnerability exploited to deploy an IP stealing attack. Hence, the same side channels could be used for \textbf{runtime monitoring} to secure the ML system against IP stealing \cite{SideChannel}. 
\section{Reliability Threats for ML systems} \label{reliability}
 
Security threats are not the only cause for an ML system not to work as expected. This section discusses several environmental/natural factors, which lead to reduced ML system reliability. 

\subsection{Hardware Faults}
Errors in the hardware components that build up a system are generally classified into transient, intermittent, and permanent faults \cite{hardError3}\cite{HWfaults}. As the name implies, \textit{transient faults} induce temporary errors in the system. \textit{Intermittent faults}, on the other hand, may cause recurring system glitches. Like transient faults, intermittent faults can be removed from the system, often by the use of additional circuitry. \textit{Permanent faults} have a lasting impact on the system, and can be removed mainly by replacing the faulty hardware component. 

\subsubsection{Transient Faults}
The nature of applications where the ML systems are deployed expose these edge devices to harsh operating conditions like high temperature and altitude. These conditions, in addition to the increasing circuit clock frequencies, voltage reduction and technology scaling, have been continuously increasing the occurrence of transient faults in systems over past decades \cite{SoftError}. Transient faults can be random, i.e., occurring unpredictably, or non-random, i.e., can be reproduced under certain circumstances \cite{hardError3}. Electrostatic discharge, electromagnetic radiation, noise in hardware interconnections, or flaw(s) in fabrication are among the leading factors contributing to transient faults \cite{ESD-error}\cite{HWfaults}. 

\begin{figure}[h]
	\centering
	\includegraphics[width=\linewidth]{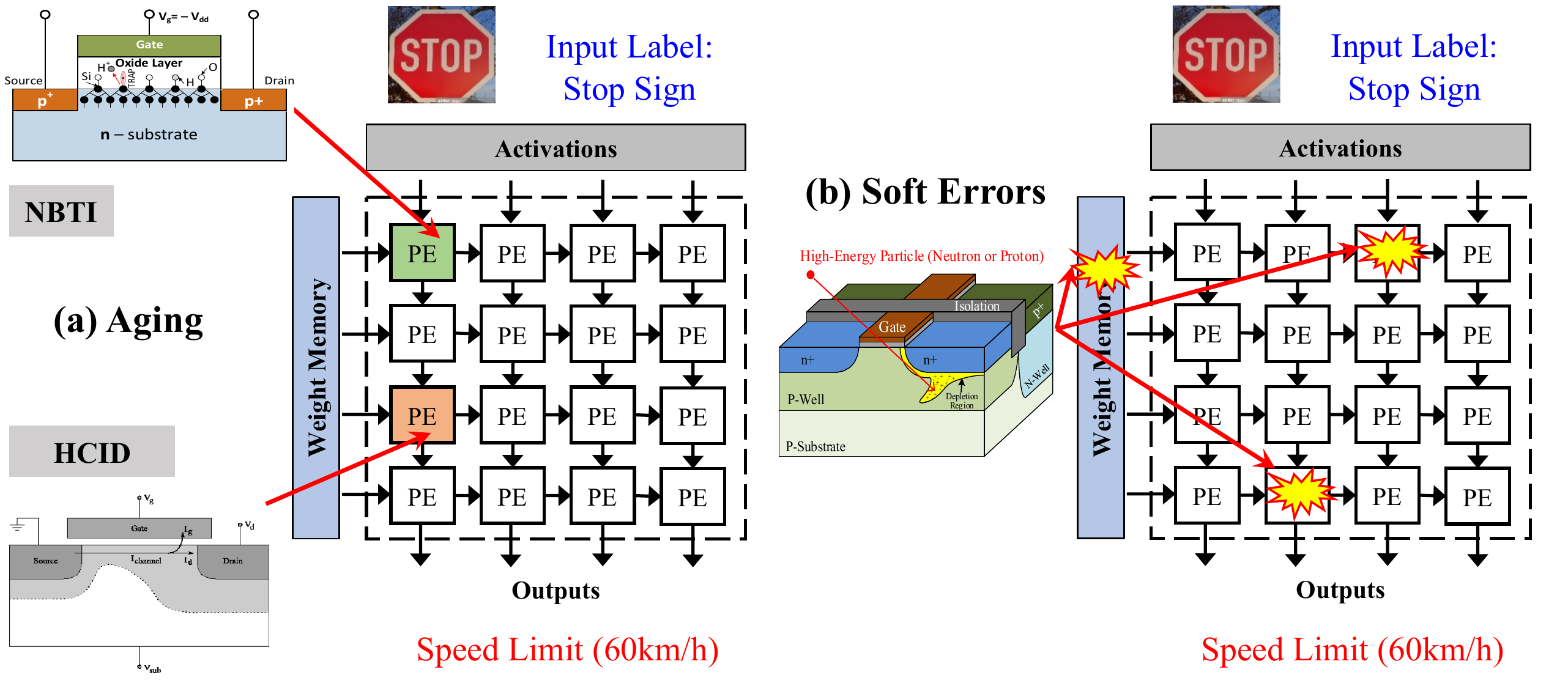}
	\caption{\textit{Effects of reliability threats, i.e.,  (a) soft errors, and (b) aging on ML systems.}}
	\label{RT}
\end{figure}

\textbf{Soft errors} \cite{SoftError} are a type of transient fault, mostly caused either by 1) a high-speed proton strike from cosmic rays, or 2) the emission of an alpha particle from impurities in IC packaging. Both particles generate a charge $Q_{rad}$ in the chip, and if this charge exceeds a certain threshold value $Q_{th}$, it is likely to change the state of the chip, resulting in a bit-flip. This effect, known as the \textit{Single Event Upset (SEU)}, is becoming a leading cause of concern with system hardware, particularly memory chips \cite{softErrorML, meza2015revisiting}. With increasing technology miniaturization, such bit-flips can extend to multiple bits within a single data word \cite{mbu-def, meza2015revisiting}. This phenomenon, termed as a \textit{Multiple Bit Upset (MBU)}, poses a challenge to robust machine learning. These effects may lead to misclassification in ML systems, as shown in Fig.~\ref{RT}(a). 

\subsubsection{Intermittent Faults}
Such faults are intermittent and  relatively unpredictable, which make them difficult to repeat, analyze, and understand.
\textbf{Process variation} \cite{kriebel_survey}\cite{processVar} is the phenomenon that results in small differences in the physical characteristics of seemingly identical circuit components during fabrication. This may lead to intermittent faults, potentially leading to permanent damage to the system chip \cite{hardError3}. Similarly, \textbf{aging} \cite{ageing} (see Fig.~\ref{RT}(b)) can cause deterioration of system performance and functions over time.
Another important factor contributing to intermittent faults in hardware is \textbf{temperature} under which the edge device is operating. Temperature effects~\cite{Temp_effect} reduce system reliability by increasing device aging and error rates. 

Often as the result of component aging, \textbf{timing errors} occur, where the system is unable to provide correct output within the expected time. Usually, as the error propagates through the chain of components, the magnitude of error increases. Not only does this reduce ML classification accuracy, it may also make the ML model vulnerable to serious security concerns~\cite{CPS_measures}. 

Accessing memory with a specific access pattern can introduce \textbf{access pattern dependent faults}, which could be caused by disturbance errors. These faults create a security vulnerability known as Rowhammer~\cite{kim2014flipping,mutlu2019rowhammer}, which is the phenomenon that repeatedly accessing a row in a modern DRAM chip causes disturbance errors in physically-adjacent rows.
DRAM data \textbf{retention failures}~\cite{khan2014efficacy,liu2013experimental,qureshi2015avatar,khan2017detecting} can also cause intermittent and unpredictable faults due to DRAM variable retention time and data pattern dependence.

\subsubsection{Permanent Faults}
These faults are irreparable, where the system portrays fixed/repetitive errors like \textit{stuck-at} faults. Factors contributing to permanent faults include \textbf{cosmic radiation}, \textbf{electrostatic discharge (ESD)} in device, \textbf{fabrication flaws}~\cite{hardError3,hardError1,ESD-error}, or recurring intermittent faults. 

\subsection{Neural Network Anomalies}\label{sec:anomalies}

\textbf{Environmental noise} ($EN$) \cite{noise-in-physical-world1}\cite{noise-in-physical-world2} has a similar impact on edge devices as adversarial attacks have on DNNs.
\begin{equation}
    f(x) \neq f(x+EN)
\end{equation}
For instance, for an object classification system, possible environmental noise could be due to fog or pollution in atmosphere, which can produce effects of blurring on the input. Similarly, variations in \textbf{data acquisition} by the edge sensors can also lead to faulty inference in an ML system. For an image acquisition system deployed in an autonomous vehicle, change in either \textit{brightness}, \textit{contrast}, \textit{camera angle}, or any other \textit{photometric transform} \cite{deepxplore}\cite{adv-camera-anlges} can impact the decision-making of the vehicle and may lead to serious consequences \cite{uber}.

\begin{figure}[ht]
	\centering
	\includegraphics[width=\linewidth]{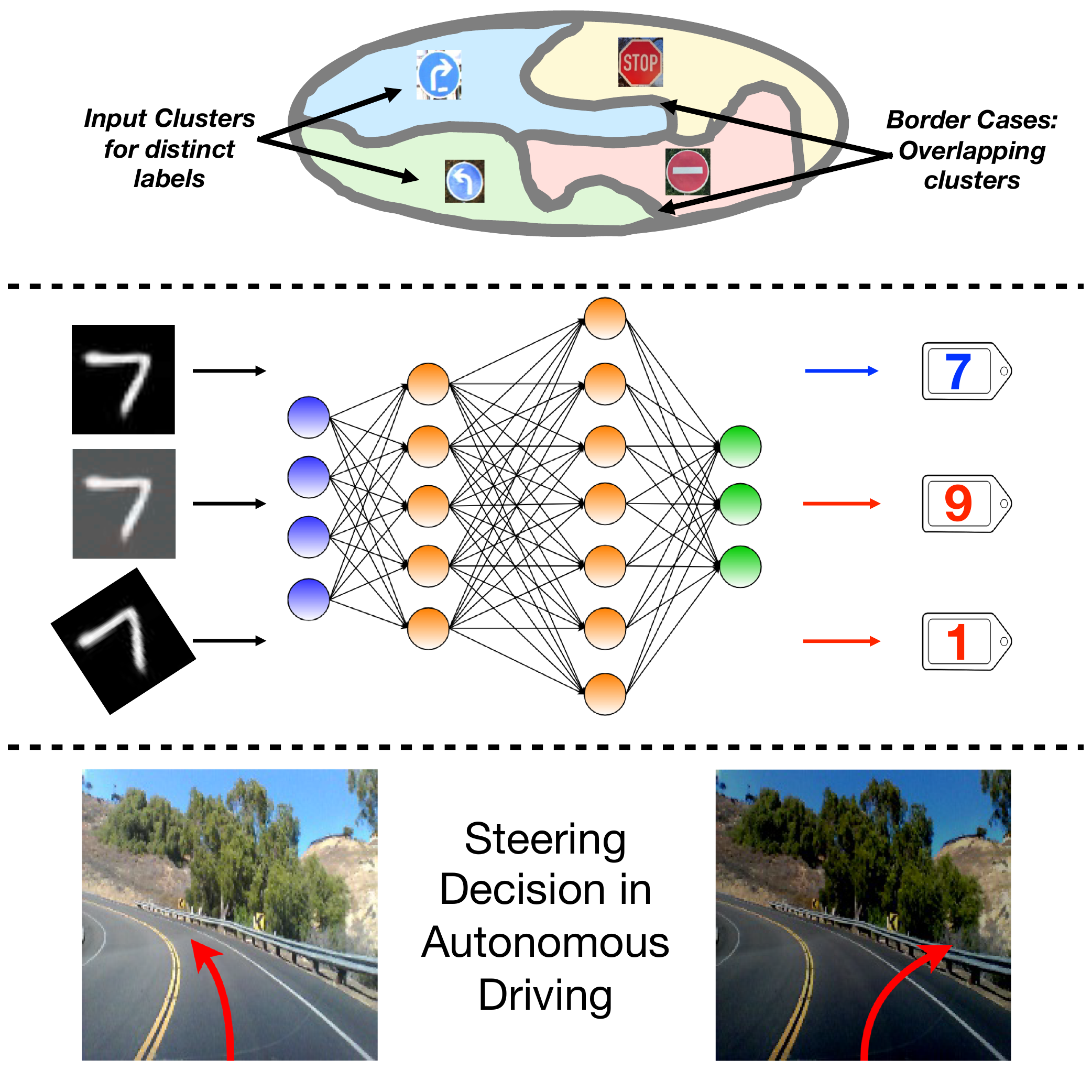}
	\caption{Inputs close to cluster boundaries (top) in hyperspace are most vulnerable to environmental adversarial transformations. Variation during data acquisition (middle) can cause misclassification, which can lead to drastic effects in ML systems (bottom) \cite{deepxplore}.}
	\label{env}
\end{figure}

The reason for such DNN anomalies is a lack of generalization of DNN for unseen inputs. The classification boundaries of the DNN outputs may overlap in the hyper-space, as depicted for a 2D space in Fig.~\ref{env} (top). The inputs closer to these boundaries are vulnerable, and slight input changes, even in the absence of a malicious attacker, may lead to misclassification.

\section{Mitigation techniques for reliability threats in ML systems}\label{sec:mitigation}

This section discusses several mitigation techniques for the reliability threats in ML systems discussed in Section~\ref{reliability}.

\subsection{Mitigation Techniques for Hardware Faults}
The most notable approaches to ensure system reliability in presence of the various hardware faults are as follows.

\subsubsection{Protection against Transient Faults}
Generally, transient faults can be removed by a \textit{component reset} or \textit{system reboot}. However, these are often not the most desirable solutions. \textbf{Interleaving} to prevent errors in consecutive bits \cite{interleaving}, using additional circuitry for \textbf{error detection} \cite{bisc}, \textbf{scrubbing} to periodically remove errors to prevent error accumulation~\cite{scrubbing, qureshi2015avatar}, adding \textbf{hardware redundancy} and voting mechanisms \cite{TMR} to rule out the erroneous bits, and using \textbf{error detection and correction codes} \cite{hamming, hsiao, SoftError, patel2019understanding}, are generally the preferred choices to defend against soft errors in memories and logic. Recently, replicating the complete hardware accelerator and conjoining the accelerators with majority voting is also being used to ensure safety in ML systems. For instance, Tesla's self-driving car computer has two chips deployed to tolerate faults~\cite{tesla2brains}.

Numerous approaches are available to handle transient errors; yet, all these approaches provide a trade-off between error detection/correction capability, area, power consumption and latency. Redundancy-based approaches can incur large area overhead and cost. A recent work shows that, in a DNN-based system, the bit flips from $1$ to $0$ have a more drastic effect on the system's classification accuracy than bit flips from $0$ to $1$ \cite{PoisoningComparison}. This finding could be used for system design with stronger error correction mechanisms deployed for more critical bit flips.

\subsubsection{Protection against Intermittent Faults}
As system components age at different rates, components in the same chip may require different levels of protection. Protection techniques that are consistent throughout the system, like chip-level guardbanding, may thus not be sufficient. A recent work \cite{ageAware} studies dynamic protection approaches that ensure that the most vulnerable components receive the highest protection in the system. The same work also proposes \textbf{age-aware workload management} to age all components of the system at the same rate. Disturbance errors like Rowhammer can be mitigated via probabilistic mechanisms~\cite{kim2014flipping} and various other hardware or software techniques~\cite{mutlu2019rowhammer}. Online profiling of memory cells~\cite{liu2013experimental,khan2014efficacy,patel2017reach,qureshi2015avatar,lee2017design,khan2016parbor} can also help the system to discover and disable weak cells with intermittent or aging-related errors.

To detect timing errors, several studies propose to use \textbf{Razor flip flops} \cite{TE-Drop}\cite{razor-approach2}\cite{voltscale_timingerror}. Once a timing error is detected, error correction is usually employed by either introducing slack in computation, skipping a clock cycle, or scaling voltage to mitigate the error's effect. However, these approaches may introduce a delay in execution as the correct result propagates to the output. Another mitigation approach to defend against timing errors is \textbf{formal timing analysis} \cite{FM-timing-analysis-1}\cite{FM-timing-analysis-2}. Such timing verification approaches are intended to ensure that the system behaves correctly within the defined timing bounds.

\subsubsection{Protection against Permanent Faults}
\begin{figure}[ht]
	\centering
	\includegraphics[width=\linewidth]{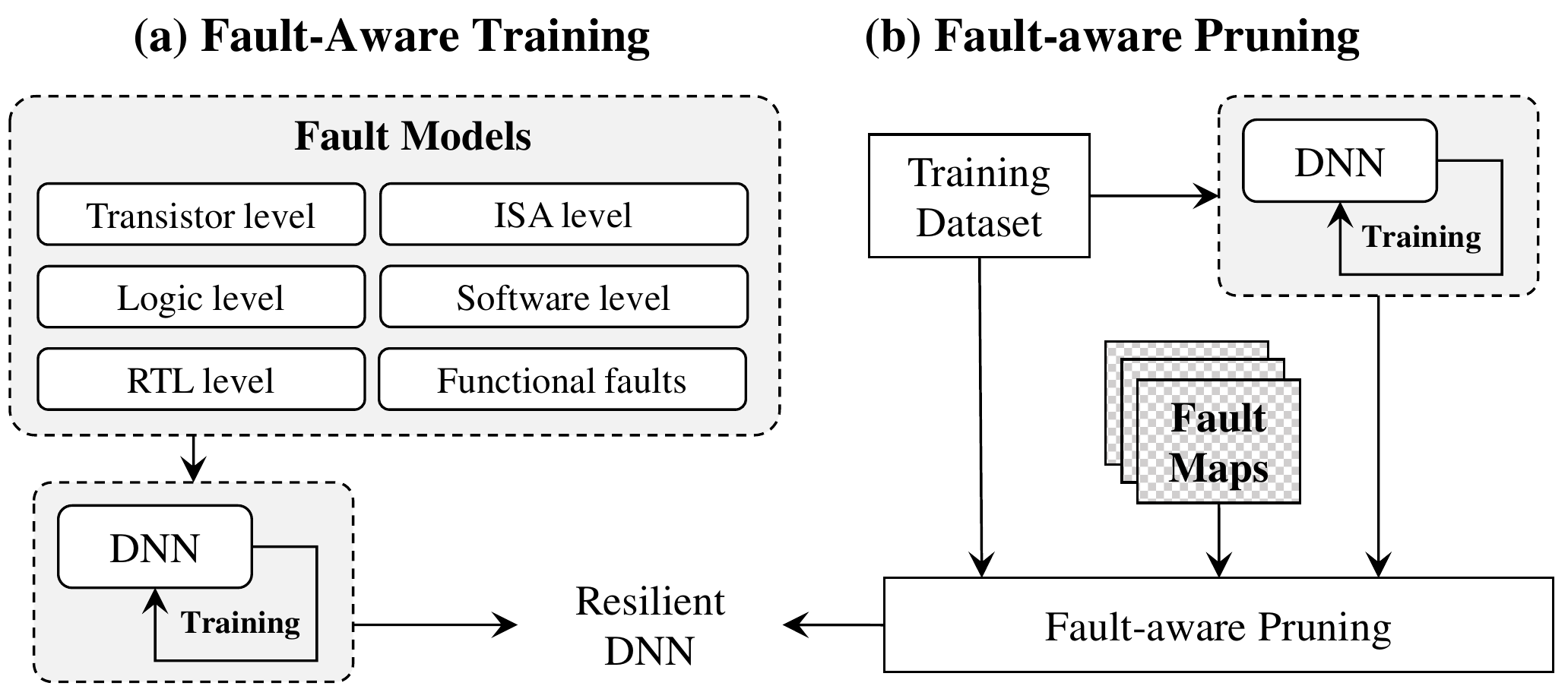}
	\caption{\textit{Mitigation techniques for permanent faults in ML systems}}
	\label{mit_permanent}
\end{figure}

Hard errors imply irreversible chip damage, for which the most effective solution is usually to \textbf{replace} the faulty chip/component. However, this is a costly solution. A relatively cost-effective alternative to chip replacement is discarding only the erroneous bits/byte of the component \cite{ECP,PAD}, which minimizes the cost incurred. Specific to ML systems, techniques like fault-aware  training, pruning, mapping and activation clipping are often used to address permanent faults~\cite{zhang2018analyzing,softErrorML,koppula2019EDEN,abdullah2020salvagednn,hoang2019ft}. In \textbf{fault-aware training}, the DNN is trained for different faults at multiple levels, like transistor level and logic level, as shown in Fig.~\ref{mit_permanent}(a). This is a computational greedy solution. In \textbf{fault-aware pruning}, all the DNN connections and parameters that map to faulty processing elements/nodes are pruned using fault maps of the baseline hardware (e.g., systolic array-based accelerator), as shown in Fig.~\ref{mit_permanent}(b). In \textbf{fault-aware mapping}, the saliency of DNN parameters is exploited to define a mapping of different segments of the DNN. This mapping is then used to prune the ineffectual parameters of the DNN while retaining the salient parameters. In \textbf{activation clipping}, the activation values exceeding the predefined threshold for a fault-free neural networks are clipped. This eliminates the need for either pruning or retraining.

\subsection{Mitigation Techniques against Environmental Noise} 
Similar to defenses against adversarial attacks (Section \ref{adv-defenses}), \textbf{pre-processing filters} \cite{fadeML} can reduce the effects of environmental noise in DNNs. Likewise, \textbf{adversarial training} \cite{retraining} of the DNN with noisy inputs could improve DNN accuracy for certain noise patterns. However, similar to the effect of using adversarial training for adversarial attacks, this solution may not work well because adversarial training overfits the network to adversarial examples but does not ensure better generalization \cite{retraining_fails}. Since the accuracy of ML systems in the presence of environmental noise and varying input data arise due to the lack of generalization to unseen inputs in the DNN, an alternative solution could be to train the DNN on a larger input dataset. However, it is not always possible to obtain a large and diverse input dataset. To overcome this limitation, some works propose the generation of \textbf{synthetic datasets}~\cite{synthetic-data}\cite{synthetic-data2}\cite{synthetic-data3}\cite{synthetic-data4}. Yet, real input domains are mostly very large, multi-dimensional, and continuous spaces. Hence, it is uncertain if any \textit{finite} number of synthetic input points could be sufficiently representative of the \textit{entire} input domain, allowing the trained DNN to generalize for unseen inputs.

\section{Formal Verification for Robust ML} \label{fm}

\begin{figure*}[ht]
	\centering
	\includegraphics[width=\linewidth]{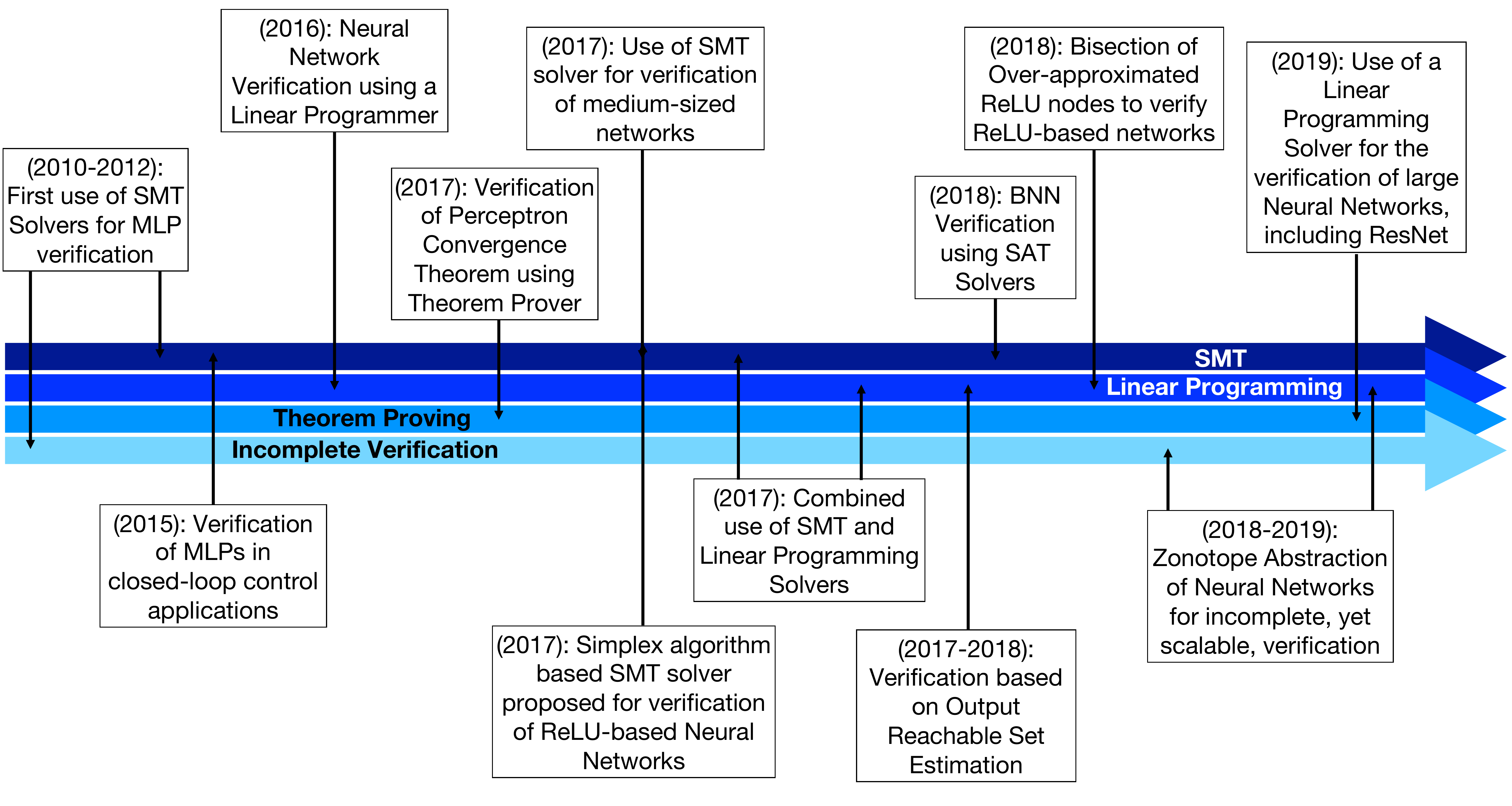}
	\caption{\textit{A decade of Verification Techniques for Neural Networks}}
	\label{timeline}
\end{figure*}
As briefly highlighted in Section \ref{taxonomy}, testing a trained DNN using a labelled dataset is insufficient to ensure reliable DNN inference. This is due to the lack of generalization of DNNs for unseen inputs. Recently, efforts have been made to understand and interpret the decision making process inside the DNNs, hoping to provide dependable guarantees regarding DNN inference. These include exploring input feature space\cite{grad-cam}, using saliency maps to understand DNN inference \cite{Intepret-saliencymaps}, and developing various certifiability criteria for DNN interpretability \cite{self-explaining-dnn}\cite{intepretability}\cite{2019certifiably}.

Formal verification provides an orthogonal alternative to testing that provides formal/mathematical guarantees regarding NN performance at the edge. The use of formal verification for hardware and software has existed for a long time \cite{FMforHW}\cite{FMforSW}. Yet, research on verification of neural networks, which forms an essential component of the ML system, has been an active domain of research for only a decade. Fig. \ref{timeline} summarizes the major milestones reached in NN verification over time, according to the four major verification categories: SAT/SMT solving, linear programming, theorem proving, and incomplete verification.

\subsection{SAT/SMT}
\begin{figure}[h]
	\centering
	\includegraphics[width=\linewidth]{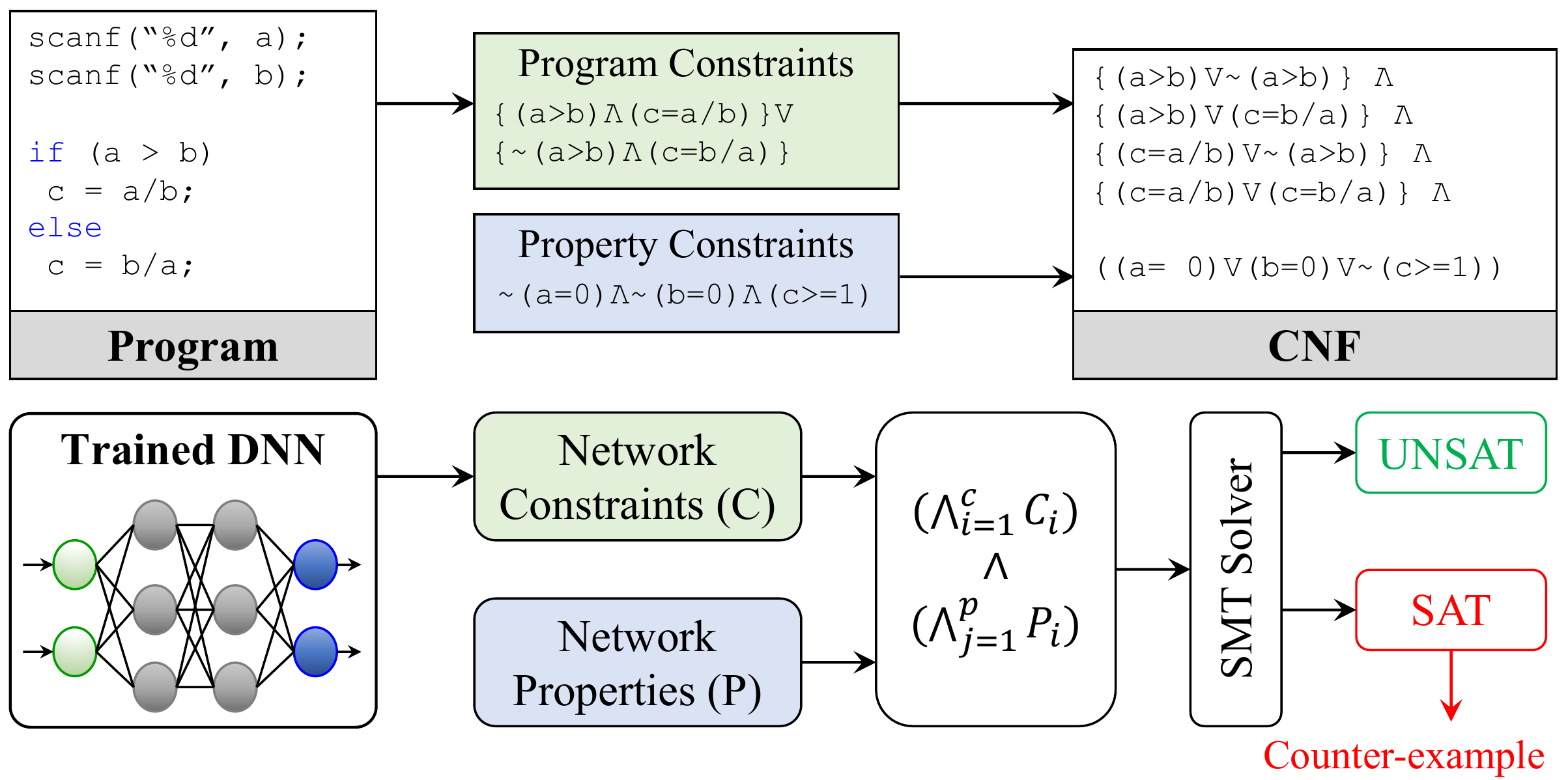}
	\caption{\textit{Using an SMT solver for verification. CNF expresses program and property constraints of C code (top), and SAT/SMT solver for a DNN-based system (bottom).}}
	\label{sat}
\end{figure}

Satisfiability (SAT) checking is the branch of formal verification where the system model and the property to be verified for the system are expressed in propositional logic, and written into Conjunctive Normal Form (CNF), as shown in Fig.~\ref{sat} (bottom). The formula is then checked by an automatic SAT solver. Having a \textit{SAT} output implies that a satisfying solution to the negation of the property, i.e., a counter-example, has been found. An \textit{UNSAT} output implies the absence of any counter-example, and hence indicates that the stated property holds for the system. Satisfiability Modulo Theories (SMT) is a variant of SAT that works similar to SAT solving, as shown in Fig.~\ref{sat} (top), but allows the use of theories beyond propositional logic, like linear arithmetic.

Since SAT solving allows the use of only propositional variables (i.e., atoms), it is often the verification approach of choice for Binarized Neural Networks (BNNs) \cite{BNNverification}\cite{FORTISS_sat}. 
SMT solvers, on the other hand, are preferred for verifying DNNs with real and/or integer network parameters \cite{reluplex}\cite{z3}. Another concept often associated with SAT-based verification approaches is Counter-Example-Guided based Abstraction Refinement (CEGAR) \cite{cegar}, which produces more reliable verification results by iteratively improving the network model using counter-examples. CEGAR and its variants provide an efficient verification solution when the DNN is modeled using over-approximation \cite{cetar}.

However, SAT-based verification suffers from the scalability problem: state-of-the-art techniques are capable of verifying only small networks \cite{pulina2012}\cite{isat3}, comprising of less than $10$ neurons, to medium-sized networks \cite{z3}, comprising of up to $20,000$ neurons. Although some works propose optimizations, like $K$-factoring \cite{FORTISS_sat}, to reduce the size of this problem, applying  these optimizations can be computationally costly. More rigorous and cost-effective optimizations can improve the scalability problem with SAT-based DNN verification.

Another challenge is to design more efficient SAT/SMT solvers. There has been a tremendous improvement in state-of-the-art SAT solvers in recent years, with increased computational speed and capability to deal with larger networks. Yet, there is a lack of dedicated tools for DNN verification; existing tools \cite{reluplex} are not scalable to larger networks. More powerful SAT/SMT solvers could be key for the improvement of DNN verification.

\subsection{Linear Programming}
\begin{figure}[ht]
	\centering
	\includegraphics[width=\linewidth]{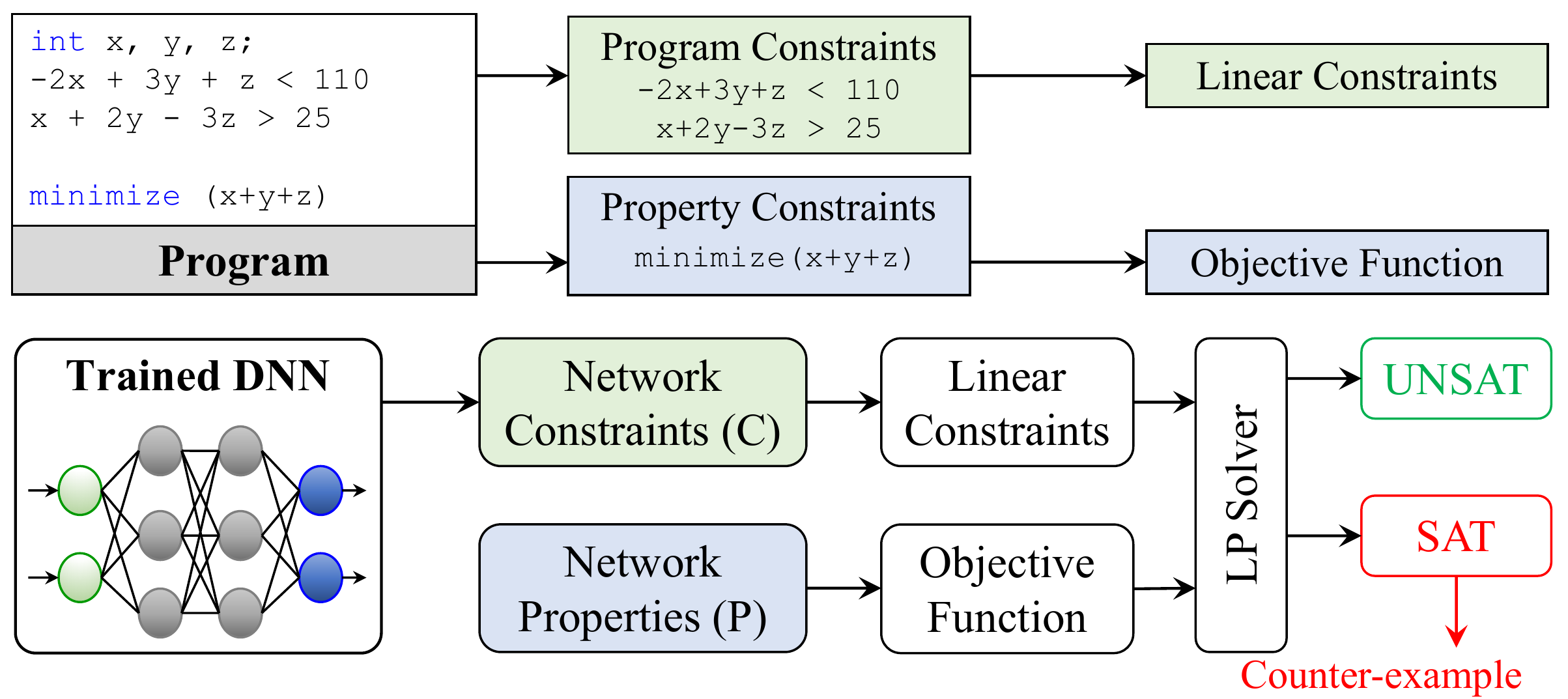}
	\caption{\textit{Using a Linear Programmer to define the linear constraints and the objective function (top), and verification of a  DNN-based system with a Linear Programmer Solver (bottom)}}
	\label{lp}
\end{figure}

Linear Programming (LP) based verification works by defining the system as a set of linear constraints, and the property to be verified as an objective function, as shown in Fig. \ref{lp}. The objective function can be either a minimization or a maximization function. The search of the minima/maxima is automatic, and involves the use of linear programmers \cite{Gurobi}\cite{cplex}.

For DNN verification, LP is generally used to check the robustness of the network against adversarial attacks. The objective is to determine the smallest noise (or noise margin) that satisfies linear constraints of the network but causes misclassification at network output \cite{bastani}\cite{MIT}.

As the name suggests, an inherent limitation of LP is that it requires the constraints to be linear. For DNNs, this poses a problem due to the presence of non-linear activation functions. Some works \cite{DeepZ}\cite{DeepPoly}, as will be discussed in Section \ref{sec-iv}, replace non-linear activation functions by their linear approximations. This yields incomplete verification results since  a linear representation is insufficient to fully replicate the behavior of the actual non-linear activation function. Another approach, proposed for ReLU-based networks, is input bisection for selected network nodes \cite{Neurify}. ReLU is a piece-wise linear function that works like a half-way rectifier: output is zero if the input is negative but output equals the input for all non-negative input values. A calculated input bisection splits ReLU into two linear functions, at the cost of a larger size verification problem.

The use of Big-M encoding\footnote{The Big-M technique is used for the verification of ReLU based networks, where a binary indicator variable $Y$ is added to the linear constraints to indicate the linear region of the activation function to which the constraint belongs, while $M$ provides a valid output upper bound that is greater than the maximum output value of every ReLU node in the network. We refer the reader to \cite{bigMtechnique} for details of the technique.} is proposed in several recent works~\cite{dutta2018,DataPoisoning,FORTISS_lp,lomuscio2017,lomuscio2018}. Although the approach ensures reliable verification results, without a significant increase in the size of the problem, it also suffers from the scalability problem. Reducing the number of constraints by eliminating the inactive neurons \cite{MIT}, and exploiting the sparsity of practical DNNs may allow effective verification of practical-sized ML systems.

\subsection{Interactive Theorem Proving}
\begin{figure}[ht]
	\centering
	\includegraphics[width=\linewidth]{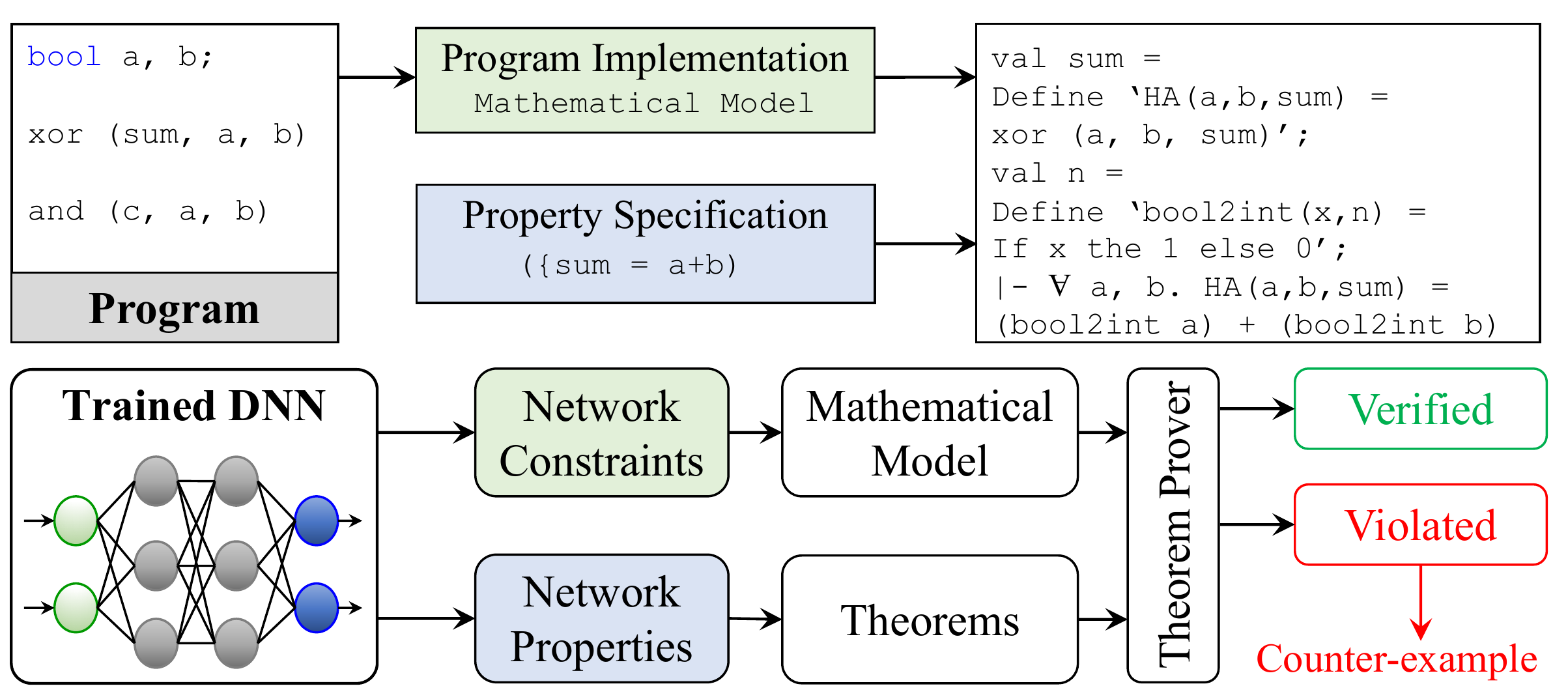}
	\caption{\textit{Using Theorem prover for verification of a half-adder (top), and mathematical model and theorems of Theorem Proving for a DNN-based system (bottom).}}
	\label{tp}
\end{figure}

Theorem proving is a type of formal verification in which the system and its properties are defined mathematically, and the properties are verified for the system by rules of natural deduction \cite{clarke1996formal}. The verification example demonstrated in Fig.~\ref{verification1} shows how natural deduction based reasoning works. Fig.~\ref{tp} gives a more generic view of how theorem proving works. Generally, for propositional logic and simple circuits, state-of-the-art theorem provers are able to verify the system without human intervention, i.e., these systems can be verified by automatic theorem provers. However, for complex systems, like DNNs, human guidance is essential, and hence verification of such systems is done via interactive theorem proving.

For verification, the system is represented as a logical model governed by mathematical principles. The property is similarly expressed as a formal proof goal. The objective is to use axioms and rules derived from these axioms to check if the properties, i.e., system specifications, hold for the system model, i.e., the implementation. 

As expected from a human-guided verification approach, interactive theorem proving is difficult to execute for two reasons. First, it requires an in-depth knowledge of the underlying system for realistic system modeling. Second, it demands the verifier to have an expert understanding of 1) why a certain property holds for the system, 2) what are the required assumptions, and 3) how to prove the property on the basis of sound mathematical principles. Hence, it is no wonder that interactive theorem proving has been a scarcely explored research domain for DNN verification. \cite{Verified_ConvTH} verifies the perceptron convergence theorem, but the work focuses on a very small subset of DNNs called binary classifiers, and hence may not be easy to adopt for large state-of-the-art DNNs.

For more practical theorem proving based DNN verification approaches, the basic need is to understand how DNNs work, why they make certain decisions, and what are the mathematical reasons behind their behavior. The perceptron convergence theorem \cite{rosenblatt1957}\cite{rosenblatt1961} was proposed almost six decades before it was formally verified by \cite{Verified_ConvTH}. Hence, understanding and developing the theory behind DNN operation seems to be  a logical step before theorem proving could be successfully employed for DNN verification.

\subsection{Incomplete Verification} \label{sec-iv}
\begin{figure}[ht]
	\centering
	\includegraphics[width=\linewidth]{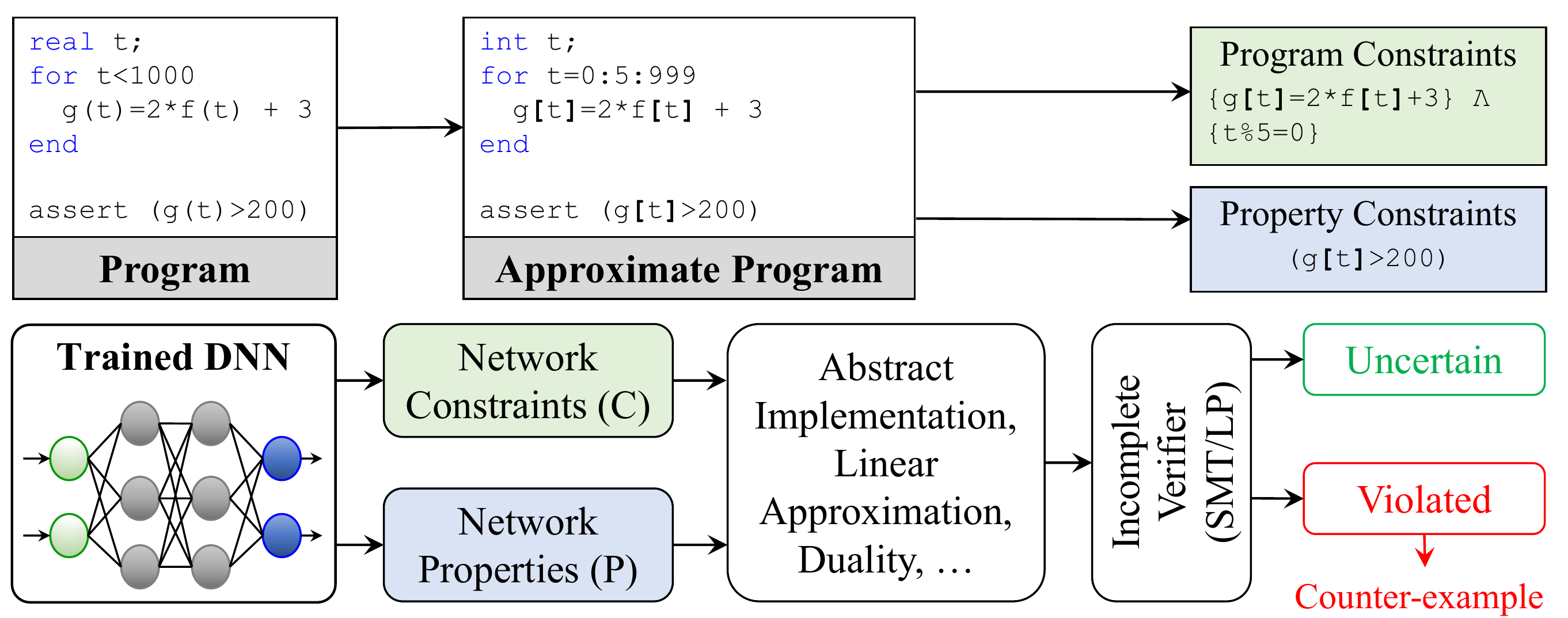}
	\caption{\textit{Using Incomplete Verification for 1) verifying a continuous-domain program (top), and 2) verifying a DNN-based system (bottom).}}
	\label{iv}
\end{figure}

Completeness is a notion that decides whether a system model is sufficient to prove everything about the system. Incomplete verification often makes use of abstract interpretation, linear approximation and other similar approaches to formally model the system \cite{ai2,matlab_safetyGuided,matlab1}. As a result, the system model is not an exact representation of the actual system but rather an over-approximation. Verification is then performed on this approximate model, as shown in Fig. \ref{iv}. It is important to note that simulation/testing, which also provides incomplete results, must not be confused with incomplete verification. This is because, in testing, the system is considered a black-box and the tester analyzes the system behavior by feeding the black-box with a finite set of inputs and recording the output. In contrast, in incomplete verification, the system is a white-box representing the simplified version of the actual system, on which formal verification is performed.

Since incomplete verification involves verifying a simplified version of the actual system model, this makes the approach scalable, even to larger DNNs \cite{DeepZ}\cite{DeepPoly}. To improve the completeness of verification, we can use abstraction refinement approaches like CEGAR \cite{cegar}\cite{cetar}. This does not entirely eliminate the problem of incompleteness of verification, but improves the reliability of verification results.

Incomplete verification often leads to false positives \cite{DeepZ}\cite{DeepPoly}. Whenever the incomplete verifier provides counter-examples, they are actual scenarios where the property does not hold for the system. If the verifier provides no counter-examples, the system may still be unsafe or the property being verified may still not hold for some inputs to the system \cite{matlab2}.

Incomplete verification is scalable, and, hence, is an attractive verification alternative for DNNs. Yet, its inherent incompleteness provides the biggest limitation to its accuracy. A possible solution is to trade off some scalability of incomplete verification with completeness~\cite{RefineAI}. This can be accomplished by iteratively refining the network model until it matches the exact system model \cite{cegar}, or combining incomplete verification with other complete verification approaches like SMT solvers or LP.
\section{Open Challenges and Discussion}\label{challenges}

Although ML is a rapidly evolving domain, it will probably pass a long time until ML systems are considered robust. In ML systems, similar to other systems, a single vulnerability is sufficient to pose a security or reliability issues that might prevent the system from obtaining accurate results. However, it is very challenging to provide strong robustness guarantees, because we need to deal with a wide range of security and reliability threats. This section discusses some important (in our view) open challenges for achieving robust ML systems.

ML systems have numerous security issues mainly related to 1) outsourced training, 2)  untrusted fabrication foundries, and 3) attacker access to the environment in which the system is deployed. Among these security issues, some of the most important are the following:
\begin{enumerate}[leftmargin=*]
    \item \textit{Securing training datasets} before outsourcing them for training. This may involve encrypting the training dataset from the cloud servers to ensure IP privacy, or minimizing the impact of data poisoning attacks.
    \item \textit{Obfuscating ML hyper-parameters, algorithms and IPs}. There are several defenses to successfully obfuscate ML hyper-parameters, algorithms and IPs using blurring and noise addition. However, as indicated earlier, these techniques do not often work well in practice. Hence, a prospective obfuscation method could be the inclusion of various obfuscation techniques in a single framework, and random switching between these techniques to ensure a more secure ML system.
    \item \textit{Ensuring fairness of training}, i.e., preventing bias of the trained NN. Gradient-based adversarial input generation and  counter-examples generated via formal verification of NNs can be used to identify bias in training. This method is based on the observation that adversarial inputs are more likely to identify the output classes to which the trained NN is biased towards. 
    \item \textit{Validating the functional and behavioral correctness of ML hardware}. Formal verification methods may be required to provide stronger guarantees on the ML hardware operation by performing verification under diverse security and reliability conditions.
    \item \textit{Minimizing the accessibility to side-channel leakages}. This can be achieved by minimizing the sharing of resources like memory and power, hence ensuring the hardware isolation of the ML system. However, this may be a costly solution for most ML applications. Another prospective solution to ensure minimal access of an attacker to the side-channel leakages can be the introduction of complementary synthetic noise to nullify the side-channel signatures of the system.
\end{enumerate}

The DNN model and the hardware that runs the model are both vulnerable to inconsistencies in performance over their life time. Major unresolved reliability challenges in ML systems include:

\begin{enumerate}[leftmargin=*]
    \item \textit{Developing frameworks to emulate ML systems} under diverse operating conditions. This is essential to 1) study and better understand the reliability challenges of the systems deployed in physical environment, 2) assess the performance of the available mitigation techniques, and 3) 
    analyze the trade-offs between these approaches to identify the solution that ensures the highest system reliability.
    \item \textit{Providing a fault-safe runtime} in case of system discrepancies. Currently, such fault-safe techniques include the use of redundancies at hardware and software levels, which ensure that, in case of a component malfunctioning, the overall performance of the system is not affected. However, these measures are generally very costly, and hence, there still exists the need for better fault-safe mechanisms for ML systems.
    \item \textit{Hampering the progression of subsystem failures} to the interconnected components. This requires mitigation approaches that can provide cross-layer reliability to ensure that a failure in one system component does not propagate and affect the results of the next system component(s). 
\end{enumerate}

Formal verification is a promising way to provide strong robustness guarantees in ML systems via mathematical proofs. The major challenges for making formal verification a practical tool to ensure robustness include:
\begin{enumerate}[leftmargin=*]
    \item \textit{Formally modeling} the non-linear, non-convex behavior of ML systems. Complete verification with existing modelling approaches (e.g., Big-M)  is often not the optimal solution due to the large number of generated clauses/constraints. Incomplete verification is also not the optimal solution because it may lead to false positive results due to over-approximation.
    \item \textit{Incorporating the uncertainties of the real world into the formal system model}. Namely, the verification of the system under different reliability factors, e.g., environmental noise.
    \item \textit{Inspecting system behavior for all possible inputs}. Formal verification is widely acclaimed due its rigorous analysis and complete results. However, due to the complexity of NN verification, current approaches rely on applying verification to only a subset of inputs (i.e., seed inputs). Providing complete guarantees regarding system behavior requires more rigorous verification approaches.
    \item \textit{Optimizing the verification goal} to reduce the computational complexity of the verification problem. As the size of the underlying ML system increases, the size of its formal representation also increases. This requires large computational overhead and time to formally verify ML systems. Hence, simplifying the verification problem prior to the actual verification can reduce the computational complexity of the problem.
    \item \textit{Improving the timing efficiency of verification}, while ensuring the completeness of verification results. There is a trade-off between the timing cost of verification and the completeness of the verification results. With the development of efficient verification tools, the bridge between timing efficiency and completeness has been reduced. However, achieving the most optimal trade-off between timing efficiency and completeness still remains an open challenge.
    \item \textit{Scaling the verification algorithm} to be applicable to practically-sized DNNs. With improvements in verification tools and formalization approaches, the size of the DNNs that can be formally verified is increasing gradually.
\end{enumerate}

Tackling the previous challenges and research directions is important for providing secure and reliable ML systems. However, as ML is a domain that advances very rapidly, there will probably be new challenges and research directions that will become important with the emergence of new ML models, deeper DNNs, unreliable hardware with reduced technology nodes, and new attack models.
\section{Conclusion}

Machine Learning (ML), particularly Neural Networks (NNs), forms an essential component of modern  Cyber-Physical Systems (CPSs). However, due to outsourced training, compromised foundries, stealthy attackers, system aging, and the harsh operating environment of these systems, both at the system cloud and edge, they are vulnerable to numerous security and reliability concerns. This survey highlight 1) the most prominent security and reliability challenges for ML systems, 2) the mitigation approaches to defend the systems against these challenges, and 3) formal methodologies for  verifying trained NNs. This survey also summarizes the most important open challenges that hamper robust ML systems.


\bibliographystyle{IEEEtran.bst}
\bibliography{ref}
\begin{IEEEbiography}[{\includegraphics[width=1in,height=1.25in,clip,keepaspectratio]{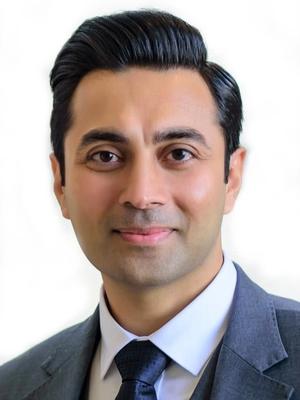}}]{Muhammad Shafique} (M'11 - SM'16) is a full professor of Computer Architecture and Robust Energy-Efficient Technologies (CARE-Tech.) at the Institute of Computer Engineering, TU Wien, Austria since Nov. 2016. He received his Ph.D. in Computer Science from Karlsruhe Institute of Technology (KIT), Germany, in Jan.2011. Before, he was with Streaming Networks Pvt. Ltd. where he was involved in research and development of video coding systems for several years. His research interests are in computer architecture, power-/energy-efficient systems, robust computing, hardware security, Brain-Inspired computing trends like Neuromorphic and Approximate Computing, hardware and system-level design for Machine Learning and AI, emerging technologies \& nanosystems, FPGAs, MPSoCs, and embedded systems. His research has a special focus on cross-layer modeling, design, and optimization of computing and memory systems, as well as their deployment in use cases from Internet-of-Things (IoT), Cyber-Physical Systems (CPS), and ICT for Development (ICT4D) domains.

Dr. Shafique has given several Keynote, Invited Talks, and Tutorials. He has also organized many special sessions at premier venues and served as the Guest Editor for IEEE Design and Test Magazine and IEEE Transactions on Sustainable Computing. He has served on the TPC of numerous prestigious IEEE/ACM conferences. Dr. Shafique received the 2015 ACM/SIGDA Outstanding New Faculty Award. six gold medals in his educational career, and several best paper awards and nominations at prestigious conferences like CODES+ISSS, DATE, DAC and ICCAD, Best Master Thesis Award, DAC'14 Designer Track Best Poster Award, IEEE Transactions of Computer "Feature Paper of the Month" Awards, and Best Lecturer Award. Dr. Shafique holds one US patent and has (co-)authored 6 Books, 10+ Book Chapters, and over 200 papers in premier journals and conferences. He is a senior member of the IEEE and IEEE Signal Processing Society (SPS), and a member of the ACM, SIGARCH, SIGDA, SIGBED, and HIPEAC.
\end{IEEEbiography}

\begin{IEEEbiography}[{\includegraphics[width=1in,height=1.25in,clip,keepaspectratio]{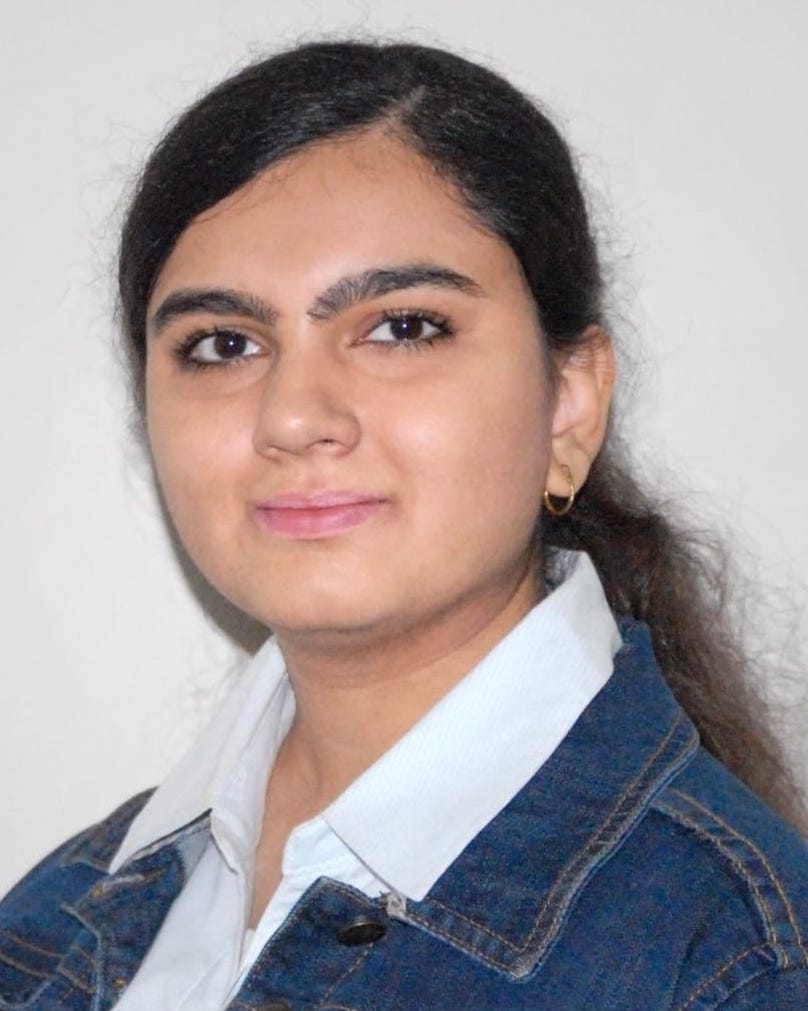}}]{Mahum Naseer}(S'19) received her B.E. in Electronics Engineering degree from NED University of Engineering and Technology, and M.S. in Electrical Engineering degree from National University of Sciences and Technology (NUST), Pakistan, in 2016 and 2018 respectively. She is currently pursuing her Ph.D. degree in Formal Methods for Resilient Embedded Systems from Technische Universit\"at Wien (TU Wien), Austria. Her research interests include reliability analysis of systems, error control coding, resilient machine learning systems, and formal methods for system verification.
\end{IEEEbiography}

\begin{IEEEbiography}[{\includegraphics[width=1in,height=1.25in,clip,keepaspectratio]{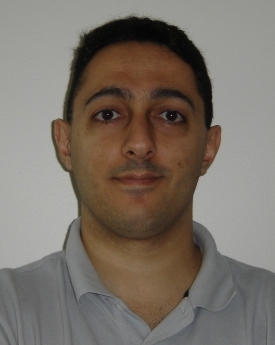}}]{Theocharis (Theo) Theocharides} (S'01-M'05-SM'11) is an Associate Professor in the Department of Electrical and Computer Engineering, at the University of Cyprus. Theocharis received his Ph.D. in Computer Engineering from Penn State University, working in the areas of low-power computer architectures and reliable system design with emphasis on computer vision and machine learning applications. Theocharis was honored with the Robert M. Owens Memorial Scholarship in May 2005. He has been with the Electrical and Computer Engineering department at the University of Cyprus since 2006, where he directs the Embedded and Application-Specific Systems-on-Chip Laboratory. He is also a Faculty Member of the KIOS Research and Innovation Center of Excellence since the Center's inception in 2008. His research focuses on the design, development, implementation and deployment of low-power and reliable on-chip application-specific architectures, low-power VLSI design, real-time embedded systems design and exploration of energy-reliability trade-offs for Systems on Chip and Embedded Systems. His focus lies on acceleration of computer vision and artificial intelligence algorithms in hardware, geared towards edge computing, and in utilizing reconfigurable hardware towards self-aware, evolvable edge computing systems. He serves on several organizing and technical program committees of various IEEE and ACM conferences, is a Senior Member of the IEEE, and a member of CEDA, is a member of the ACM, and an Associate Editor for IEEE Consumer Elect ronics magazine, the IET Computers and Digital Techniques, and the ETRI journal. He also serves on the Editorial Boards of IEEE Design \& Test magazine, and ACM Transactions on Embedded Computing Systems.
\end{IEEEbiography}

\begin{IEEEbiography}[{\includegraphics[width=1in,height=1.25in,clip,keepaspectratio]{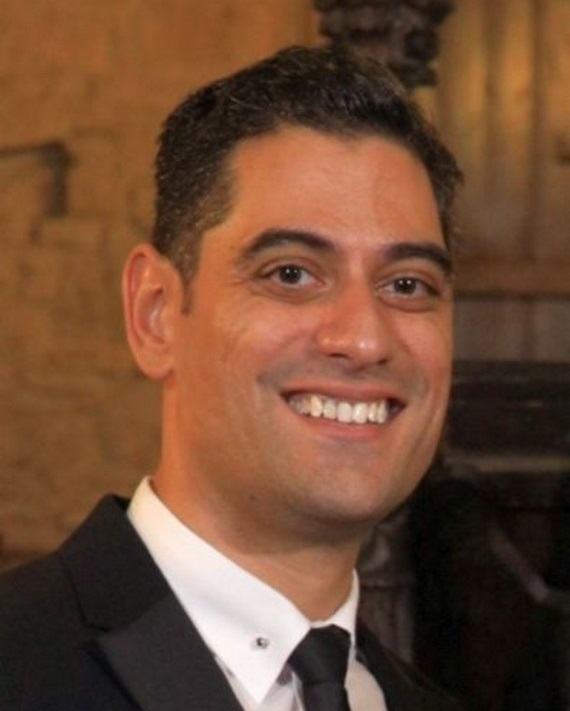}}]{Christos Kyrkou} (S'09-M'14) received the B.Sc., M.Sc., and Ph.D. degrees in computer engineering from University of Cyprus, Nicosia, Cyprus, in 2008, 2010, and 2014, respectively. He is a Research Associate with the KIOS Research Center for Intelligent Systems and Networks, University of Cyprus. His research interests include real-time embedded systems, fieldprogrammable gate arrays and reconfigurable hardware, computer vision, machine learning, and smart camera networks.

Dr. Kyrkou is a member of ACM and the Technical Chamber of Cyprus. He is a Technical Program Committee Member of the IEEE International Symposium on Nanoelectronic and Information Systems and the International Conference on Pervasive and Embedded Computing. He received an award for graduating top of his class during his B.Sc. studies at University of Cyprus and a full scholarship for his postgraduate studies from the Department of Electrical and Computer Engineering, University of Cyprus.
\end{IEEEbiography}

\begin{IEEEbiography}[{\includegraphics[width=1in,height=1.25in,clip,keepaspectratio]{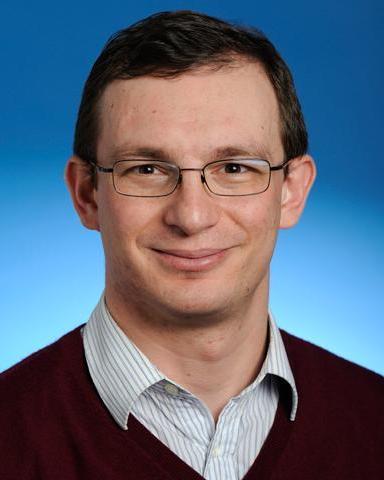}}]{Prof. Onur Mutlu} is a Professor of Computer Science at ETH Z{\"u}rich. He is also a faculty member at Carnegie Mellon University, where he previously held the William D. and Nancy W. Strecker Early Career Professorship. His current broader research interests are in computer architecture, computing systems, hardware security, robust systems, and bioinformatics. He is especially interested in interactions across domains and between applications, system software, compilers, and microarchitecture, with a major current focus on memory and storage systems. A variety of techniques he, together with his group and collaborators, have invented over the years have influenced industry and have been employed in commercial microprocessors and memory/storage systems. He obtained his PhD and MS in ECE from the University of Texas at Austin and BS degrees in Computer Engineering and Psychology from the University of Michigan, Ann Arbor. His industrial experience spans starting the Computer Architecture Group at Microsoft Research (2006-2009), and various product and research positions at Intel Corporation, Advanced Micro Devices, VMware, and Google. He was the recipient of the ACM SIGARCH Maurice Wilkes Award, the inaugural IEEE Computer Society Young Computer Architect Award, the inaugural Intel Early Career Faculty Award, faculty partnership awards from various companies, a healthy number of best paper or ”Top Pick” paper recognitions at various computer systems and architecture venues. He is an ACM Fellow, an IEEE Fellow "for contributions to computer architecture research and practice", and an elected member of the Academy of Europe (Academia Europaea). His computer architecture course lectures and materials are freely available on YouTube, and his research group makes software artifacts freely available online. For more information, please see his webpage at http://people.inf.ethz.ch/omutlu/.
\end{IEEEbiography}

\begin{IEEEbiography}[{\includegraphics[width=1in,height=1.25in,clip,keepaspectratio]{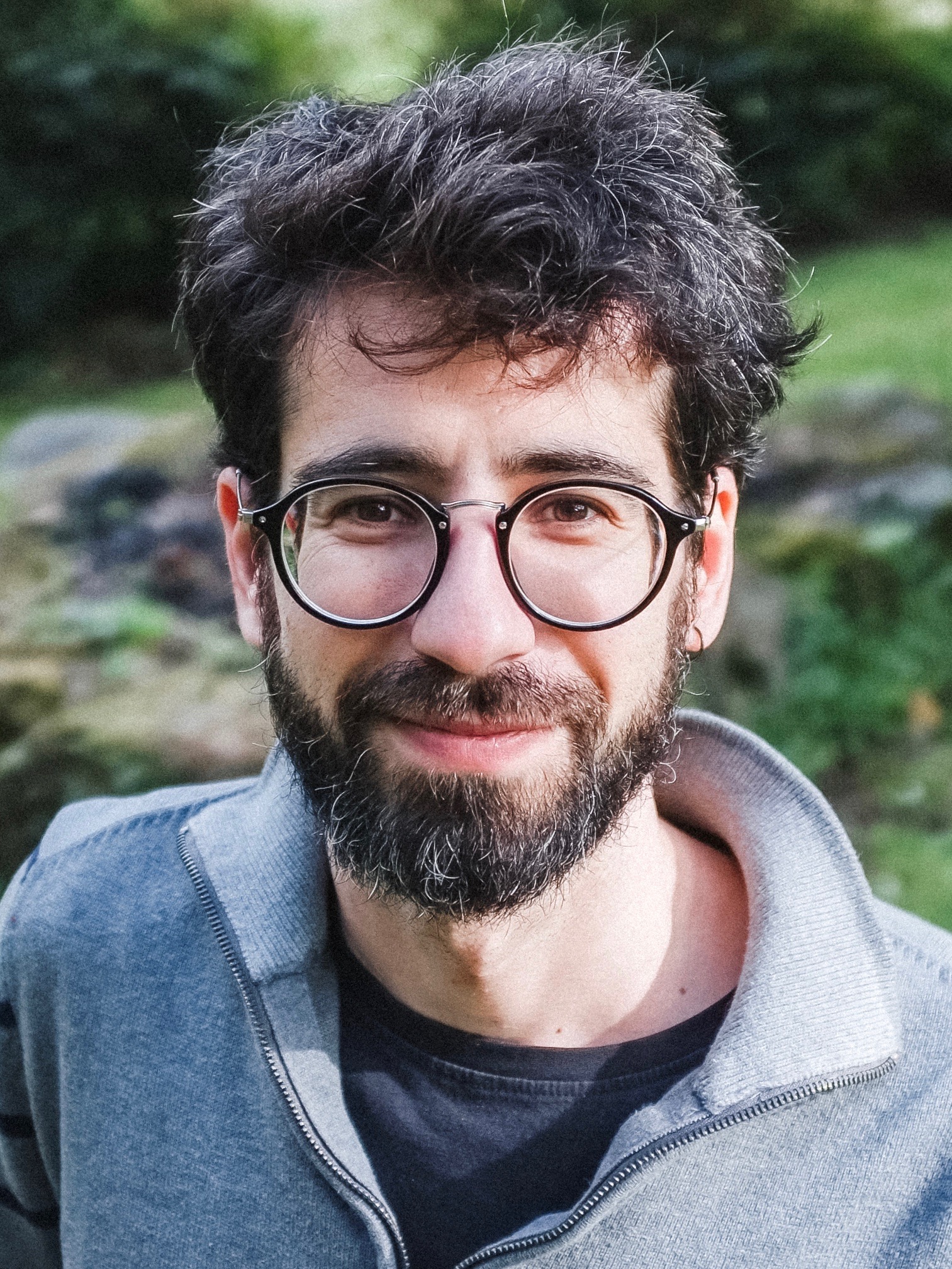}}]{Lois Orosa} is a PostDoc in the SAFARI research group at ETH Z{\"u}rich. His current research interests are in computer architecture, hardware security, memory systems, and ML accelerators. He obtained his PhD from the University of Santiago de Compostela, and he was a PostDoc in the Institute of Computing at University of Campinas. He was a visiting scholar at University of Illinois at Urbana-Champaign and Universidade NOVA de Lisboa, and he acquired industrial experience at IBM, Recore Systems, and Xilinx.
\end{IEEEbiography}

\begin{IEEEbiography}[{\includegraphics[width=1in,height=1.25in,clip,keepaspectratio]{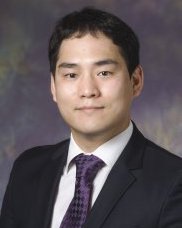}}]{Jungwook Choi} is currently an assistant professor at Hanyang University, South Korea. He was formerly  a research staff member at IBM T.J. Watson Research, Yorktown Heights, NY, where he has worked in the areas of approximate machine learning and deep learning algorithms for hardware acceleration. His research interests include high performance, energy efficient, and reliable implementation of machine learning and deep learning algorithms. He has a PhD in electrical and computer engineering from the University of Illinois at Urbana-Champaign.
\end{IEEEbiography}

\section*{}\label{Appendx}
\begin{table*}[h] 
  \centering
  \resizebox{.98\linewidth}{!}{

  \begin{tabular}{m{3cm}  m{5cm}  c }
 
    \multicolumn{3}{l}{\normalsize APPENDIX}\\
    \\ \\
    
    \hline
    \textbf{Neural Network (NN)} & \textbf{Description} & \textbf{Pictorial Representation of the Network} \\ \hline
    
    Feed-Forward Neural Network
    &
    These are the neural networks with neurons in every layer impacting only the decision of neurons in the successive layers. Hence, the networks are cycle/loop - free. The feed-forward networks are also called \textit{fully-connected} when every neuron in the preceding layer is connected to every neuron in the successive layer.
    & 
    \begin{minipage}{0.5\textwidth}
      \includegraphics[width=\linewidth]{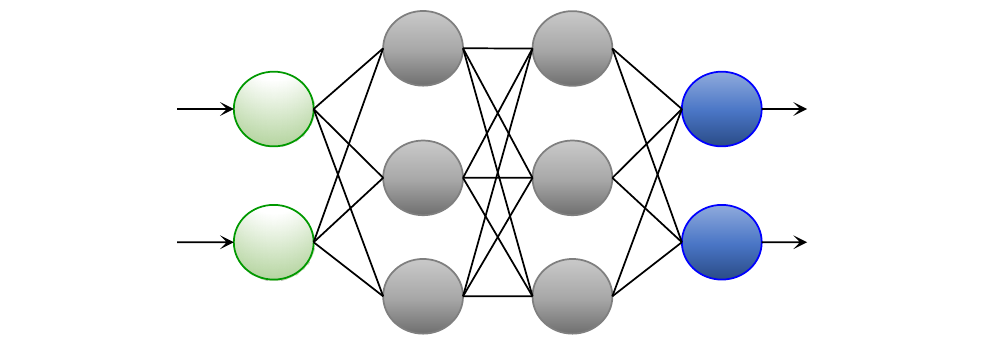}
    \end{minipage}
    \\ \hline
    
    Recurrent Neural Network (RNN)
    &
    RNNs comprise of feedback loop(s); hence, neurons in one layer can impact the values of neurons in successive as well as preceding layers. This provides temporal characteristics to the RNNs, i.e., the values of the neurons (or the internal memory of the network) varies temporally.
    & 
    \begin{minipage}{0.5\textwidth}
      \includegraphics[width=\linewidth]{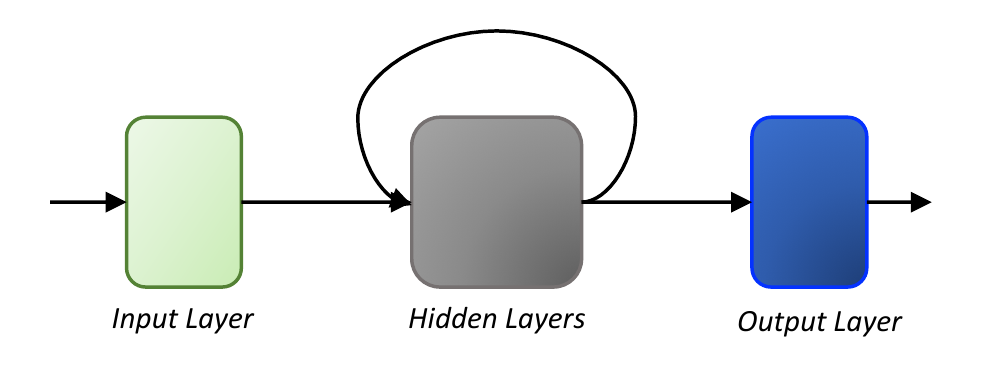}
    \end{minipage}
    \\ \hline
    
    Convolutional Neural Network (CNN)
    &
    Unlike the earlier fully-connected networks, CNNs share network weights via convolution operation. This improves the local spatial correlation of the input, and ensures that only the most prominent input features of the input are carried to the successive network layers.
    & 
    \begin{minipage}{0.5\textwidth}
      \includegraphics[width=\linewidth]{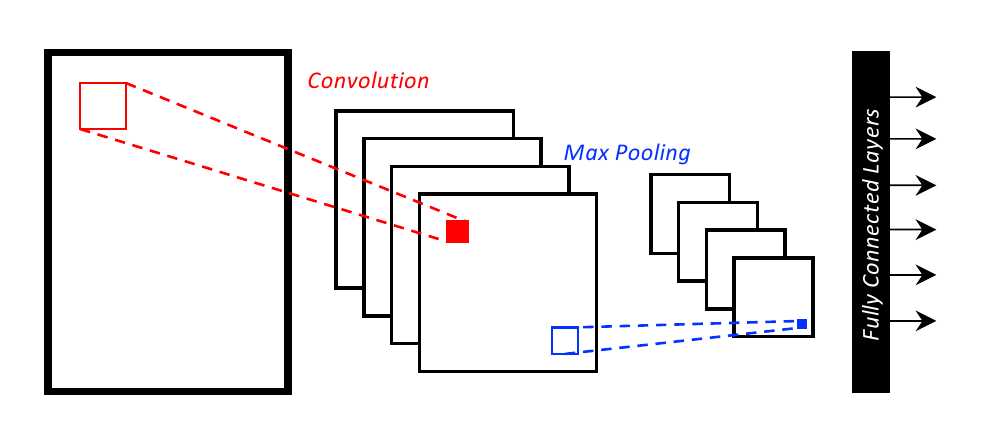}
    \end{minipage}
    \\ \hline
    
    Generative Adversarial Network (GAN)
    &
    GANs involve an interplay between a generator and a discriminator for the training of the network. The generator produces synthetic inputs in the same latent space as the training dataset, while the discriminator learns to distinguish the original data from the synthetic data. Hence, the objective of the generator is to maximize the error (i.e., generate more realistic synthetic inputs) while the discriminator minimizes the error by learning to differentiate between real and synthetic input.
    & 
    \begin{minipage}{0.5\textwidth}
      \includegraphics[width=\linewidth]{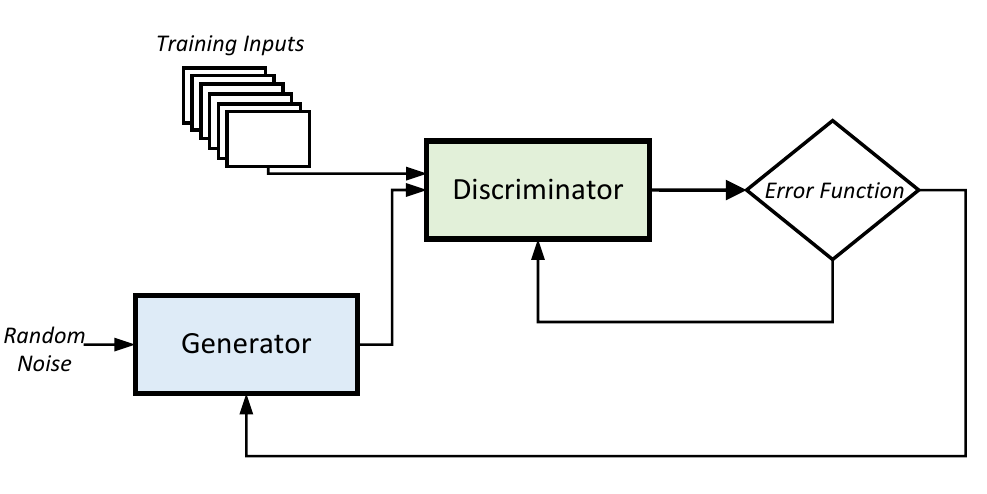}
    \end{minipage}
    \\ \hline

    Capsule Network (CapsNet)
    &
    CapsNets are build up of layers that operate on vectors, where each element of the vector represents the instantiation parameter that deduces whether the feature represented in the vector is actually present in the input. The length of the vector, on other hand, represents the instantiation probability. The connections between two consecutive capsule layers are learned dynamically during inference through the \textit{routing-by-agreement} algorithm, which iteratively updates the \textit{coupling coefficients} of the CapsNet. In this way, capsules learn to interpret high level features in a hierarchical manner.
    & 
    \begin{minipage}{0.5\textwidth}
      \includegraphics[width=\linewidth]{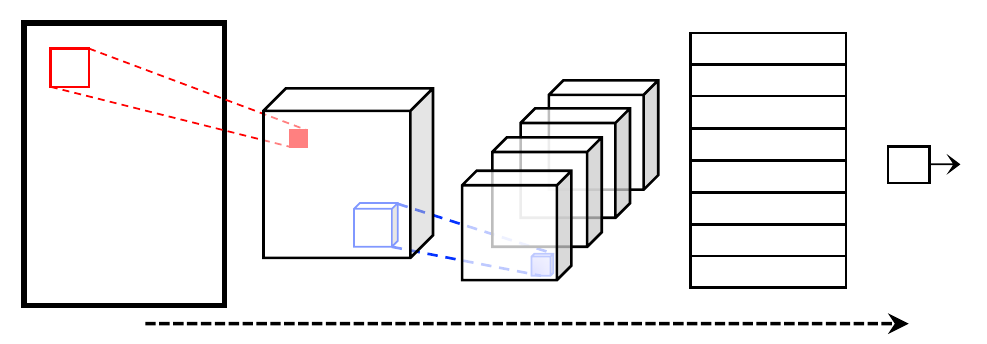}
    \end{minipage}
    \\ \hline
    
    Spiking Neural Network
    &
    All the NNs discussed above assume a normalized firing frequency for the neurons. This neglects the dynamic behavior of the inputs like speech. SNNs make use of spike trains to depict the spatio-temporal characteristics of the input. Hence, SNNs are an important class of NNs particularly for time-dependent applications. 
    & 
    \begin{minipage}{0.5\textwidth}
      \includegraphics[width=\linewidth]{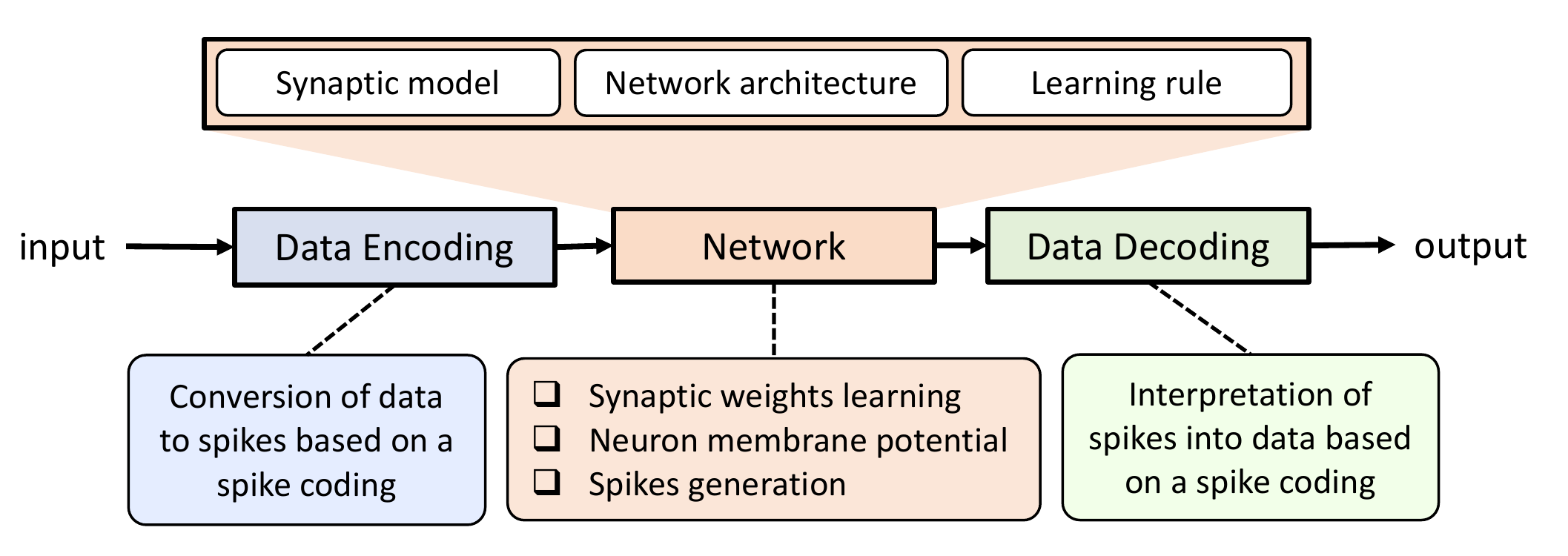}
    \end{minipage}
    \\ \hline
    
  \end{tabular}
  }
\end{table*}

\end{document}